\documentclass{aa}  
\usepackage{txfonts}
\usepackage[]{hyperref}
\usepackage{graphicx}	
\usepackage{amsmath}	
\usepackage{float}
\usepackage{color}
\usepackage{soul}
\usepackage{caption}
\usepackage{subfig}
\usepackage{comment}

\sethlcolor{yellow}

\newcommand{\G}{{\it Gaia }}

\newcommand{\mycomment}[1]{}
\usepackage{multicol}

\begin{document}

   \title{Hot DQs, magnetic and metal-polluted white dwarfs: spectroscopic insights from a \emph{Gaia} machine-learning--selected 500 pc sample}

   \subtitle{}

   \author{Enrique Miguel Garc\'{\i}a-Zamora
          \inst{1}\fnmsep\thanks{Email;
    enrique.miguel.garcia@upc.edu}
          \and
          Santiago Torres\inst{1,2}
          \and
          Alberto Rebassa-Mansergas\inst{1,2}
           \and
          Aina Ferrer-Burjachs\inst{1}
                  }

\institute{Departament de F\'\i sica, 
           Universitat Polit\`ecnica de Catalunya, 
           c/Esteve Terrades 5, 
           08860 Castelldefels, 
           Spain
           \and
           Institut d'Estudis Espacials de Catalunya, Esteve Terrades, 1, Edifici RDIT, Campus PMT-UPC, 08860 Castelldefels, Spain}

\date{\today}
\titlerunning{Hot DQs, magnetic and metal-polluted white dwarfs in a \emph{Gaia} sample}
\authorrunning{Garc\'{\i}a-Zamora et al.}

\offprints{E. M. García Zamora}
 
  \abstract
   {The latest \G data release provides astrometric measurements for 1.8 billion sources and low-resolution spectra for 220 million, including approximately 100\,000 white dwarfs. Although useful for pre-classification, \G spectra lack the resolution required for accurate spectral typing and reliable parameter determination, motivating dedicated spectroscopic follow-up observations.}
  {We assess the reliability of machine-learning spectral classifications derived from \G spectra through comparison with medium-resolution spectroscopy. We determine the nature of objects classified as “massive helium-rich (DB)” white dwarfs by automated methods, and characterise the properties of warm and hot DQ (carbon-dominated) white dwarfs, as well as magnetic and metal-polluted objects.}
   {We observed 255 white dwarfs along the \G B and Q branches with the Gran Telescopio Canarias equipped with the OSIRIS instrument ($R\simeq1000$). Spectral types were assigned through visual inspection and compared with machine-learning classifications applied to \G spectra. For objects labelled as “massive DBs”, we independently determined their atmospheric spectral composition. Magnetic white dwarfs were identified via Zeeman splitting, and first-order magnetic field strengths were estimated when possible.}
   {Machine-learning classifications are highly accurate (>90\% for spectral types included in their training sets), despite the low resolution of \G spectra. We determine the true nature of objects classified as “massive DBs” to be largely composed of magnetic white dwarfs and warm DQs, with only 5 of 112 observed objects (4.46\%) confirmed as genuine DBs. Warm DQs are found along the \G Q branch and exhibit unusually high tangential velocities.}
 {We provide spectral classifications for 255 white dwarfs and demonstrate that Random Forest algorithms reliably classify low-resolution \G spectra for the main spectral types. We also determine the nature of objects classified as “massive DBs” and identify a large population of magnetic white dwarfs and carbon-rich objects. Several rare subtypes are identified, including 2 DAB, 1 DBAZ, 1 DZAB, 2 DZBA, 14 DAH, 1 DAQ, 1 DQZA, 4 hot and 29 warm DQ stars, along with 63 magnetic white dwarfs. The location and kinematics of warm DQs are consistent with previous studies, supporting their proposed origin as merger remnants.}

   \maketitle
%

\section{Introduction}
\label{s:intro}

White dwarfs, the degenerate cores of stars with initial masses lower than 8-10$M_{\odot}$, are the most common stellar remnant in the Galaxy \citep{Althaus2010}. Spectroscopic observations allow us to obtain information of their atmospheric composition, so that they can be classified into different spectral types and subtypes, following the most prominent spectral features. As such, a white dwarf that shows Balmer lines will be classified as a DA, another one showing \ion{He}{I} features will be classified as a DB; and a spectrum with metallic lines as well as magnetic Zeeman splitting will be classified as a DZH white dwarf. For a  more detailed description on white dwarf classification, see \cite{Sion1983}. 

An accurate spectral classification of white dwarfs is not only essential for accurate white dwarf parameter determination \citep{Bergeron2019,Tremblay2019b}, but also for understanding white dwarf spectral evolution (see \citealt{Bedard2024} and references therein). Furthermore, features such as the high ratio of DQs (white dwarfs displaying carbon features) inside the Q-branch \citep{Tremblay2019}, or the presence of carbon in hydrogen-deficient atmospheres as a possible explanation for the \G colour-magnitude bifurcation \citep{Camisassa2023,Blouin2023} depend on correct spectral identification to be understood. This is also true for processes such as convective mixing or dilution in white dwarfs \citep{Blouin2019, Cunningham2020}. Metal-polluted DZ white dwarfs, in turn, allow us to probe into the origin and composition of the accreted material, as determined by their abundances \citep{Zuckerman2007,Farihi2010}.

In recent years, the \G mission \citep{GaiaMission2016} has sparked a revolution in the astronomical community by providing astrometric and photometric measurements for 1.8 billion sources, as well as spectral information for 220 million objects \citep{DeAngeli2022}, including more than 100,000 white dwarfs. The spectrophotometric data, however, have a modest resolving power (30$\lesssim R \lesssim$100; \citealt{Carrasco2021}), which limits their diagnostic capability. Despite the remarkable success of automated classification techniques applied to these data (e.g. \citealt{Garcia2023,Vincent2023,Vincent2024,Perez-Couto2024,Garcia2025}), the reliability of such classifications for rare or spectroscopically complex subtypes has been found to be much lower and more uncertain, as these classes are typically underrepresented in training sets \citep{Garcia2023,Garcia2025}. In particular, objects located along the \G B and Q branches may present atmospheric compositions or even magnetic features that are difficult to disentangle at \G resolution, potentially leading to misclassifications that propagate into population studies.

Motivated by the need to assess the robustness of automated classifications and to clarify the nature of peculiar subgroups -- such as the massive objects classified as DB in \cite{Garcia2025} and \cite{Vincent2024} -- we conducted a series of spectroscopic follow-up campaigns using the Gran Telescopio Canarias (GTC). Its 10.4\,m aperture enables efficient spectroscopy of intrinsically faint white dwarfs with moderate exposure times. We selected 321 white dwarfs within 500\,pc of the Sun as targets, of which 255 were observed and subsequently analysed. The selected sample is  well suited to test classification reliability and to investigate peculiar subtypes in critical regions of the \G colour--magnitude diagram.

This paper is organised as follows. In Section~\ref{s:Description} we describe the selection of the white dwarf sample and the details of the observational campaigns. Section~\ref{s:Visual_clas} presents the visual spectral classification of the targets based on the medium-resolution spectroscopy. In Section~\ref{s:Gen_ana} we perform a general analysis of the sample, including a comparison with machine-learning classifications derived from \G spectra, as well as a detailed discussion of objects along the Q branch, the DQ population, and magnetic white dwarfs. Finally, in Section~\ref{s:Conc} we summarise our main results and outline future perspectives.

\section{The white dwarf sample and description of the observing campaigns}
\label{s:Description}

Since the 2023B observing semester at the GTC (September 2023-February 2024), we have conducted four observational campaigns in semesters 2023B, 2024B, 2025A and 2025B with different scopes and aims. In all cases, we limited the declination (DEC $>-15\,$ deg) and apparent magnitude ($G\leq19.2$\,mag) to avoid excessively low sky altitudes and excessively long exposure times, respectively. A log of the observations is provided in Table~\ref{tab:Log}.

\begin{table}
    \caption{Log of the observations}
    \label{tab:Log}
\begin{center}
    \begin{tabular}{lcc}
            \noalign{\smallskip}
            \hline\hline
            \noalign{\smallskip}
        Campaign & Intended Objects & Observed objects\\
            \noalign{\smallskip}
            \hline
            \noalign{\smallskip}
        2023B & 82 & 76 \\
            \noalign{\smallskip}
            \hline
            \noalign{\smallskip}
        2024B & 101 & 98 \\
            \noalign{\smallskip}
            \hline
            \noalign{\smallskip}
        2025A & 100 & 43 \\
            \noalign{\smallskip}
            \hline
            \noalign{\smallskip}
        2025B & 38 & 38 \\
            \noalign{\smallskip}
            \hline
            \hline
            \noalign{\smallskip}
        Total & 321 & 255 \\
            \noalign{\smallskip}
            \hline
            \end{tabular}
    \tablefoot{Number of intended and observed objects for each observing campaign at the GTC.}
\end{center}
\end{table}

\begin{table}
    \caption{Objects already reported in recent papers}
    \label{tab:Newvsold}
\begin{center}
    \begin{tabular}{lc}
            \noalign{\smallskip}
            \hline\hline
            \noalign{\smallskip}
        Reference & Objects\\
            \noalign{\smallskip}
            \hline
            \noalign{\smallskip}
        This work, (new) & 213 \\
            \noalign{\smallskip}
            \hline
            \noalign{\smallskip}
        \cite{Jewett2024} & 18 \\
            \noalign{\smallskip}
            \hline
            \noalign{\smallskip}
        \cite{Kilic2025} & 7 \\
            \noalign{\smallskip}
            \hline
            \noalign{\smallskip}
        \cite{Kilic2026} & 24 \\
            \noalign{\smallskip}
            \hline
            \noalign{\smallskip}
        \cite{OuldRouis2026} & 5 \\
            \noalign{\smallskip}
            \hline
        \end{tabular}
    \tablefoot{Number of new objects identified in this work, along with those already reported in recent studies.}
\end{center}
\end{table}

The 2023B and 2024B campaigns aimed at ascertaining the spectral classification obtained through the Random Forest algorithms described in \cite{Garcia2023,Garcia2025}. More specifically, the high precision (true positives to all positives ratio) obtained for the DB, DQ, DO and DZ spectral types, as well as the correct classification of the three magnetic DAH white dwarfs found in these two studies. For this reason, all the selected targets belonged to these pre-classified spectral types. A total of 183 objects were intended for observations during these two semesters (see Table~\ref{tab:Log}). The comparison between the predicted and observational types for these objects will allow us to test the capabilities of automatic classification algorithms (see Section\,\ref{ss:Ascert}).

On the other hand, the 2025A and 2025B campaigns focused on uncovering the true nature of a subpopulation of approximately 350 white dwarfs classified as DB by the machine-learning algorithms of \cite{Garcia2025, Vincent2024}. We refer to these objects as the `massive DB' subpopulation since they are located below the $M\geq0.95\,$M$_{\odot}$ DB cooling track of \citet{Camisassa2019} in the \G Hertzsprung-Russell (HR) diagram, on and above the Q branch, which we define as the region limited by the  0.83 M$_{\odot}$ and 1.29 M$_{\odot}$ DA cooling tracks from \cite{Camisassa2019}, and the CO-core crystallization onset (top track) and 80\% mass crystallization (bottom track) curves from \cite{Camisasasa2024}. It is noteworthy that this subpopulation has not been identified in other spectroscopic works \citep{Genest2019,Bergeron2011} nor in the volume-limited \G samples of \cite{Hollands2018}, \cite{Garcia2023} and \cite{Obrien2023}. A total of 138 objects from the \cite{Garcia2025} and the \cite{Vincent2024} catalogues were selected for observations (see Table~\ref{tab:Log}). A detailed account of these findings are given in Sect.~\ref{ss:True} and Table~\ref{tab:MDB_breakdown}. 

\begin{figure}
\centering
    \includegraphics[width=1.0\columnwidth,trim=-20 0 0 0, clip]{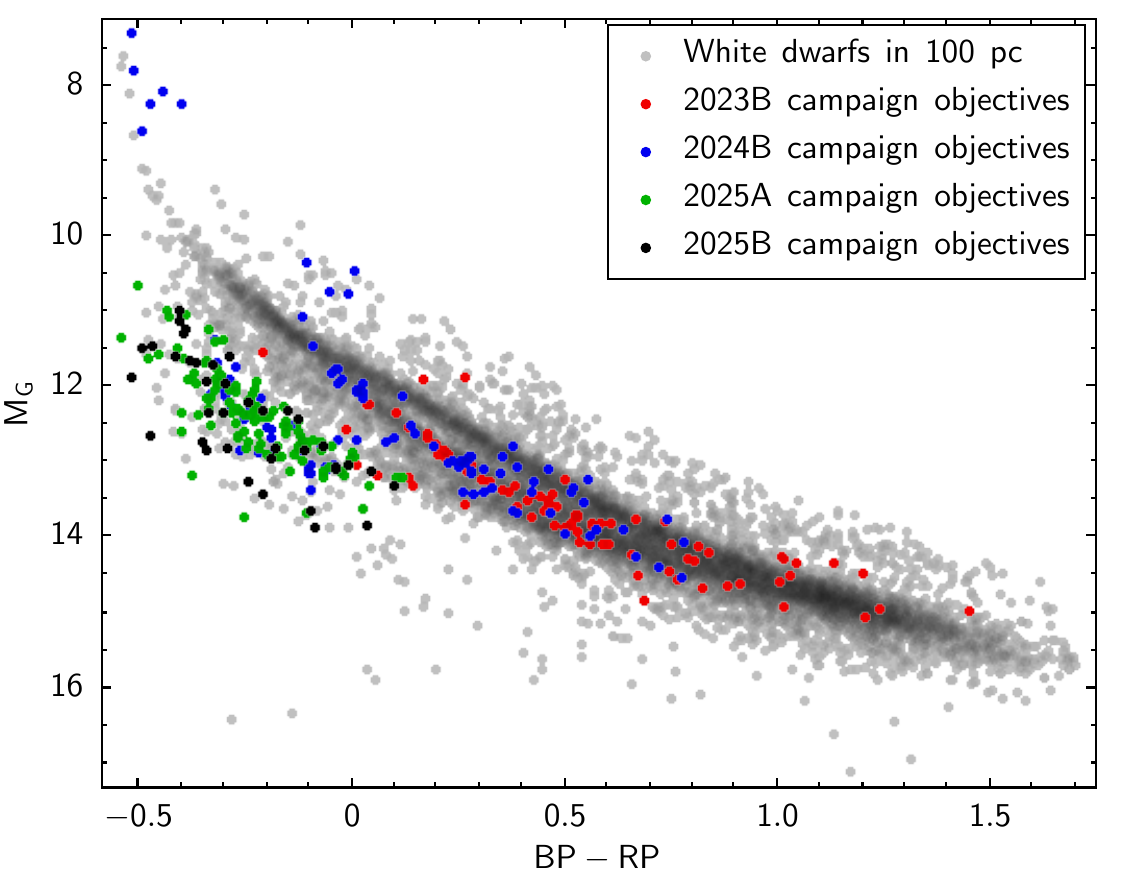}
    \caption{{\em Gaia} HR diagram showing the targets observed at the GTC during the 2023B, 2024B, 2025A and 2025B semesters, superimposed on the {\em Gaia} 100\,pc white dwarf sample of \citet{Jimenez2023}.}
    \label{f:Objectives}
        \vspace{0.5cm}
\end{figure}

Of the 321 intended objects, 255 (79.44\%) were observed (Table~\ref{tab:Log}). Of these, 213 are new, while the remaining 35 have been observed in recent papers (see Table~\ref{tab:Newvsold}). The observations were performed at any lunar brightness, a maximum seeing of 1.5" and a maximum airmass of 1.5. All four campaigns requested long-slit spectroscopy with the Optical System for Imaging and low-Intermediate-Resolution Integrated Spectroscopy (OSIRIS, see \citealt{Cepa2013}) instrument equipped with the R1000B grism and a 0.6" slit-width. This resulted in medium resolution spectra (R$\sim$1000) covering the $3600\lesssim\lambda\lesssim7800\,\AA$ wavelength range. Under the requested sky and atmospheric conditions outlined above, exposure times were defined to achieve a signal-to-noise ratio (SNR) of $\sim$35 at H$\beta$ for each spectrum in exposures of less than one hour. However, given that the observations were performed in service mode, such conditions were not always kept, which resulted in lower SNR spectra in some cases. In any case, the achieved SNR made it possible to distinguish the spectral features of the observed white dwarfs for a correct spectral classification. The GTC spectra were reduced using the \texttt{pamela} \citep{Marsh2014} and \texttt{molly} \citep{Marsh2019}\footnote{Developed by Tom Marsh and available at \url{https://cygnus.astro.warwick.ac.uk/phsaap/software/molly/html/INDEX.html}.} packages.

The {\em Gaia} HR diagram of the observed targets from each campaign, superimposed on the 100\,pc Gaia white dwarf sample from \cite{Jimenez2023}, is shown in Fig.~\ref{f:Objectives}.

\section{Visual spectral classification}
\label{s:Visual_clas}

In this section, we present a breakdown of the spectral types and subtypes assigned through visual inspection of the GTC spectra. This includes both predicted and unexpected subtype characteristics, as well as classifications that were not included in the initial training sets and therefore could not be predicted by the algorithm, such as cataclysmic variables (CVs). Selected spectra of objects described in this section can be consulted in Appendix~\ref{s:Sample_spectra}. An excerpt from the complete observed sample can be consulted in Table~\ref{tab:Muestra_catalogo}, while the full catalogue is available at CDS.

\begin{table*}[h!]
    \caption{Excerpt from the observed sample.}
    \label{tab:Muestra_catalogo}
\begin{center}
    \begin{tabular}{cccccccc}
            \noalign{\smallskip}
            \noalign{\smallskip}
Name & \textit{Gaia} source ID & RA & DEC & Predicted & Observational & Observational & \textit{B}\\
 & & (deg) & (deg) & spectral type & spectral type & spectral subtype & MG\\
\noalign{\smallskip}
\hline
\noalign{\smallskip}
J0001+5812 & 422726589831800576 & 0.33 & 58.20 & DZ & DZ & & \\              
J0006+0257 & 2739231150982741760 & 1.51 & 2.96 & DZ & DZ & & \\             
J0006+3104 & 2861452348130844160 & 1.66 & 31.07 & Massive DB & Magnetic & & \\        
J0012+5708 & 422378456967056640 & 3.18 & 57.15 & Massive DB & DC & & \\              
J0016+5251 & 419201830430458624 & 4.08 & 52.86 & Massive DB & DA & & \\              
J0019+4248 & 385166486649719552 & 4.97 & 42.81 & DZ & DZ & & \\              
J0034+7130 & 536425025677394688 & 8.64 & 71.52 & Massive DB & hot DQ & hot DQH & < 1\\
J0042+5405 & 417878327668855680 & 10.71 & 54.09 & Massive DB & DC & & \\              
J0043-1000 & 2377863773908424448 & 10.94 & -10.01 & Massive DB & DAH & & 18.7$\pm$0.4\\             
J0050+3138 & 360858960322547968 & 12.58 & 31.65 & Massive DB & DA & DAB & \\   
\end{tabular}
\tablefoot{Full Table 3 is available at the CDS.}
\end{center}
\end{table*}

\subsection{DAs}
\label{ss:DAs}

In total, we identify 26 DA white dwarfs. Of these, 3 are DAH candidates correctly classified as such by our Random Forest algorithm, 11 are DAHs classified as `massive DB' white dwarfs, and the remaining 12 are DAs misclassified as non-DAs. The misclassification of these latter objects was unexpected, as the DA/non-DA validation tests in \cite{Garcia2023,Garcia2025} showed very good performance.

\paragraph{Pure DAs}
\label{sss:DAs}
Nine of our targets are identified as pure DAs, displaying only Balmer lines. These include J0239+3733, J0313-0131, J0641+2018 (classified as DZs and a DQ, respectively), as well as J0016+5251, J0628+0622, J0726-0912, J1405-0932, J1643+6627, and J1810+1240, which were flagged as belonging to the `massive DB' subpopulation. An example spectrum can be consulted in Appendix~\ref{s:Sample_spectra}, Fig.~\ref{f:Sp1}.

\paragraph{DABs}
\label{sss:DABs}
Objects J0050+3138 and J1747+1701 display Balmer lines as well as helium features, namely the \ion{He}{I} $\lambda\lambda=4471,\,5876\,\AA$ spectral lines. Their spectra (see Fig.~\ref{f:Sp1} in Appendix~\ref{s:Sample_spectra}) are therefore classified as the DAB subtype.

\paragraph{DAHs}
\label{sss:DAHs}

Fourteen of the identified DAs show magnetic Zeeman splitting, which is specially evident around the H$\alpha$ and H$\beta$ lines. They belong, therefore, to the DAH subtype. An example spectrum can be found in Fig.~\ref{f:Sp1} in Appendix~\ref{s:Sample_spectra}; while the strength of their magnetic fields is estimated in Sect.~\ref{sss:DAH_analysis}.

\paragraph{DAQs}
\label{sss:DAQs}

DAQs are a recently discovered \citep{Hollands2020}, extremely rare (with only 26 known objects) white dwarfs, with spectra displaying Balmer and \ion{C}{I} lines. They are characterised by their high masses and peculiar kinematics \citep{Hollands2020,Kilic2024,Kilic2025}, properties that are shared with warm and hot DQs (see Sect.~\ref{ss:DQ_analysis} for a discussion and analysis of DQs, DQAs, warm and hot DQs). This is because, in reality, they belong to the same subpopulation. As shown by \cite{Kilic2024}, warm DQs display a range of hydrogen atmospheric abundances: DAQs form the H-rich end, while DQAs and DQs are found at the H-deficient end (see also Fig.~7  of \citealt{Kilic2025} and Fig.~28 of \citealt{Kilic2026}).

We visually identified one DAQ, J0325+2540. This object, also classified as such by \cite{Kilic2025}, is a hot white dwarf found in the Q branch. Its spectrum can be consulted in Fig.~\ref{f:Sp1} in Appendix~\ref{s:Sample_spectra}.

\subsection{DBs}
\label{ss:DBs}

Among the objects observationally classified as DB, we identify 9 pure DBs, 8 DBAs and 1 DBAZ. Most of these objects are not part of the `massive DB' subpopulation. In fact, only five of them are (3 DBs and 2 DBAs).

Two example spectra of a massive DB and a massive DBA are shown in Appendix~\ref{s:Sample_spectra}, Fig.~\ref{f:Sp1}.

\paragraph{Pure DBs}
\label{sss:DBs}

Nine pure DBs have been found as part of our observational campaigns. Most of them are located in the $-0.10\leq G_{\rm BP}-G_{\rm RP} \leq 0.01$ colour range (Fig.~\ref{f:HRs_analysis} in Appendix~\ref{s:HR_diagrams}, top right panel). The exceptions are the objects J0622+6935, J0903-1311 and J1832+0856, which are located in the $-0.40\leq G_{\rm BP}-G_{\rm RP} \leq-0.30$ colour range. Incidentally, these objects are part of the `massive DB' subpopulation.

\paragraph{DBAs}
\label{sss:DBAs}

We have identified eight DBAs, with both \ion{He}{I} and Balmer lines of lesser intensity. Of these, one also displays the H and K \ion{Ca}{II} lines and is treated in the following subsection. All white dwarfs are in the $-0.07\leq G_{\rm BP}-G_{\rm RP} \leq0.03$ colour range except J0646+1151, with $G_{\rm BP}-G_{\rm RP}=-0.11$. Unlike in Sect.~\ref{sss:DBs}, this outlier does not belong to the `massive DB' subpopulation.

\paragraph{DBAZ}
\label{sss: DBAZ}

Among all the observationally classified DB white dwarfs in our sample, we also find what seems to be a rare case (only 39 such objects are listed in the Montreal White Dwarf Database\footnote{\url{https://www.montrealwhitedwarfdatabase.org/}} (MWDD, see \citealt{Dufour2017})) of a DBAZ white dwarf. This object, J0201+6221, displays not only helium and hydrogen features, but also the metallic lines of the \ion{Ca}{II} H-K lines doublet, in its spectrum. This spectrum is shown in Appendix~\ref{s:Sample_spectra}, Fig.~\ref{f:Sp2}.

\subsection{DCs}
\label{ss: DCs}

In total, we find 27 DC white dwarfs. Of these, three are cool DCs ($0.43\leq G_{\rm BP}-G_{\rm RP}$), whereas the remaining 24 are hot DCs ($0.10\geq G_{\rm BP}-G_{\rm RP}$). A spectrum of a hot DC is shown in Appendix~\ref{s:Sample_spectra}, Fig.~\ref{f:Sp2}.

\subsection{DOs}
\label{ss:DOs}

We visually identify six DO white dwarfs. All of them are very hot objects ($G_{\rm BP}-G_{\rm RP}<-0.40$), and display the characteristic \ion{He}{II} lines that define this spectral type. Additionally, all of them lack \ion{He}{I} lines, except for J0513+1147, which displays prominent absorptions at $\lambda\lambda=4471,\,5876\,\AA$, and J0536+5448, which displays a weak \ion{He}{I} $\lambda=5876\,\AA$ line.

\paragraph{Pure DOs}
\label{sss:DOs}

Of the six identified DOs, two of them (J0513+1147 and J0536+5448) display exclusively helium spectral features. Therefore, they are classified as pure DOs. Incidentally, these are the only objects to display \ion{He}{I} lines. One such spectrum can be consulted in Appendix~\ref{s:Sample_spectra}, Fig.~\ref{f:Sp2}.

\paragraph{DOZs}
\label{sss:DOZs}

Four of our DOs, J0447+0458, J0515+0728, J0519-0443, and J0602-1351, show, apart from \ion{He}{II} spectral lines, \ion{C}{IV} features most prominently around $\lambda\lambda\lambda=4441,\,4658,\,5470\,\AA$, as well as a lack of \ion{He}{I} lines. Such objects should be classified as DOQs. However, due to historical reasons (see Footnote 3 in \cite{Bedard2022b}), they are classified as DOZs. We follow this convention, and assign the DOZ type to these objects. One of these spectra can be seen in Appendix~\ref{s:Sample_spectra}, Fig.~\ref{f:Sp2}.

\subsection{DQs}
\label{ss:DQs}

Among the 47 visually identified DQs, we find two distinct groups. On the one hand, there is the cool DQ subpopulation ($T_{\rm eff} \lesssim 10\,000\,$K), which displays molecular C$_{2}$ Swan bands in its spectra. On the other hand, there are the warm and hot DQ subpopulations ($T_{\rm eff}\gtrsim 10\,000\,$K), whose spectra are dominated by atomic lines of neutral \ion{C}{I} and ionised \ion{C}{II}, respectively. The limit between hot and warm DQs is not well defined, as the strength of the \ion{C}{I} and \ion{C}{II} spectral lines varies gradually with temperature rather than appearing and disappearing abruptly. \citet{Koester2019} arbitrarily sets $T_{\rm eff}=18\,000\,$K as the temperature limit between hot and warm DQs. 

\paragraph{Cool DQs}
\label{sss:coolDQ}

There are 15 cool DQs in our sample. Of these, three (J0506+2030, J0722+6143 and J1942+3402) belong to the peculiar DQ subtype (DQpec, cool DQs in which the Swan bands appear slightly blue-shifted and with a more rounded appearance than in regular DQs; see \citealt{Hall2008} for further information). All three had been correctly predicted as DQpecs in \cite{Garcia2023}. Example spectra of a cool DQ and a DQpec can be found in Appendix~\ref{s:Sample_spectra}, Fig.~\ref{f:Sp2}.

\paragraph{Warm DQs}
\label{sss:warmDQ}

We find a total of 28 warm DQs. Of these, 11 are pure warm DQs, displaying only carbon features, whereas 16 are warm DQAs that also display Balmer lines. One of the objects is a warm DQZA, with neutral oxygen features in addition to carbon and hydrogen. In all cases, we note the absence of \ion{He}{I} lines, in agreement with the results of \cite{Kilic2025}. Sample spectra of warm DQ and DQA white dwarfs can be found in Fig.~\ref{f:Sp3} in Appendix~\ref{s:Sample_spectra}. 

\paragraph{Hot DQs}
\label{sss:hotDQ}

Of the 170 `massive DB' candidates, we visually identify 4 hot DQ candidates: J0034+7130, J0420+6450, J0446+7227, and J0642+0632. All four objects also display signatures of Zeeman splitting, and are therefore classified as magnetic hot DQHs. A sample spectrum is shown in Appendix~\ref{s:Sample_spectra}, Fig.~\ref{f:Sp3}.

\paragraph{DQZA}
\label{sss:DQZA}

Finally, in object J0328+5806 we identify, in addition to \ion{C}{i} and Balmer lines, \ion{O}{i} spectral features, which are especially evident at wavelengths $\lambda\lambda\lambda\lambda = 5330,\,5436,\,6158,\,6456\,\AA$. The spectrum of this object is shown in Appendix~\ref{s:Sample_spectra}, Fig.~\ref{f:Sp3}.

Warm DQs with oxygen features are extremely rare objects. A search in the available literature reveals that \citet{Liebert2003} report two such cases, as do \citet{Gaensicke2010} and \citet{Kepler2015}, while \citet{Kilic2025} identify another new instance, as well as a doubtful case.The spectrum of this object is shown in Appendix~\ref{s:Sample_spectra}, Fig.~\ref{f:Sp3}.

\subsection{DZs}
\label{ss:DZs}

In total, we find 90 observed spectra of metal-polluted white dwarfs. The metals displayed, however, are varied, ranging from spectra showing only the \ion{Ca}{II} H and K lines to those also displaying spectral lines from iron, magnesium or sodium. 

\paragraph{Pure DZs}
\label{sss:DZs}

Of the 90 DZs, 73 (81.1\%) 74 (82.2\%) do not display additional subtype characteristics, showing only metallic lines. Fig.~\ref{f:Sp3} in Appendix~\ref{s:Sample_spectra} displays a pure DZ example spectrum.

\paragraph{DZAs}
\label{sss:DZA}

Seven other white dwarfs (J0410+0847, J0505+0048, J0543+3936, J0628+6636, J0709-1332, J1803+1634, and J2015+4742), in addition to metals, also show less intense hydrogen spectral features, in most cases only the H$\alpha$ line. This combination places them in the DZA subtype. One example is shown in Appendix~\ref{s:Sample_spectra}, Fig.~\ref{f:Sp4}.

\paragraph{DZAB}
\label{sss:DZAB}

The spectrum of J0452-0214 shows, in addition to metallic lines and hydrogen features, \ion{He}{I} spectral lines. Since the hydrogen lines are more intense than the helium lines, we classify it as the rare DZAB subtype, which only two such objects registered in the MWDD. Its spectrum can be seen in Fig.~\ref{f:Sp4} in Appendix~\ref{s:Sample_spectra}.

\paragraph{DZBA}
\label{sss:DZBA}

Furthermore, we identify two additional objects, J0437+0051 and J0745+5551, that, like J0452$-$0214, show metallic lines, as well as \ion{H}{i} and \ion{He}{ii} features. However, in these cases the He lines are more intense. We therefore classify them as the DZBA type, which has only another member (albeit a magnetic DZBAH) registered in the MWDD. The spectrum of J0745+5551 can be consulted in Appendix~\ref{s:Sample_spectra}, Fig.~~\ref{f:Sp4}.

\paragraph{DZHs}
\label{sss:DZH}

We also find six white dwarfs (J0342+2934, J0610-1402, J0738+7424, J0829-0818, J1543-0247, and J2357+2747) that show signatures of magnetic Zeeman splitting, particularly pronounced around the \ion{Na}{I} doublet at $\lambda\lambda=5890,\,5895\,\AA$, which becomes a triplet; the \ion{Mg}{I} triplet at $\lambda\lambda\lambda=5167,\,5173,\,5184\,\AA$, which splits into four components; and the \ion{Ca}{I} line at $\lambda=4226\,\AA$, which acquires a complex structure. These magnetic splittings are also present in the spectra of the first discovered DZH white dwarf \citep{Reid2001}. Therefore, we place these objects in the magnetic DZH category. For an example spectrum of a magnetic DZH, see Appendix~\ref{s:Sample_spectra}, Fig.~\ref{f:Sp4}.

\subsection{Magnetic white dwarfs.}
\label{ss:Magnetic_WD}

In total, we identify 63 magnetic objects, excluding the hot DCs described in Sect.~\ref{ss: DCs}. Of these, 14 correspond to the DAHs described in Sect.~\ref{sss:DAHs}, four to the hot DQHs described in Sect.~\ref{sss:hotDQ}, and six to the DZHs described in Sect.~\ref{sss:DZH}.

The remaining 39 objects, 36 of which belong to the `massive DB' population according to the Random Forest algorithm, span a wide range of morphologies, from a series of spectral lines with clear Zeeman splitting in J0557+1406 and J0336+0213, to one or two broad absorption bands with a width of several hundred $\AA$, with no other distinguishable spectral features. The first case is illustrated in the fifth panel of Fig.~\ref{f:Sp4}, Appendix~\ref{s:Sample_spectra}. So far, their spectral types remain uncertain, although we suspect that some of them may be DBHs. For the moment, they are grouped under the umbrella term `unclassified magnetic objects' (UMO).

\subsection{Cataclysmic Variables}
\label{ss:CVs}

CVs are binary stars composed of a white dwarf accreting from a non-degenerate companion, objects that may experience sudden and dramatic variations in their brightness. The term covers different types, both in origin and in the magnitude of the energy released, such as novae, dwarf novae and novalikes \citep[e.g.][]{Warner1995}. When the accreted material undergoes nuclear fusion, the brightness of the system increases by several orders of magnitude. 

This accreted material forms an accretion disc around the white dwarf owing to its angular momentum \citep{Williams1983}. The disc produces a two-component spectrum: a thermal continuum originating in its optically thick region and emission lines formed in the outer regions, which are optically thin in the continuum and optically thick in the lines \citep{Williams1980}. For this reason, CV spectra are characterised by the presence of H and He emission lines.

This spectral type was not included in the training set of any of our Random Forest algorithms and therefore could not be recognised as such by our classifier. We nevertheless identify two CVs through visual inspection in our GTC spectra: J2007+1742 and J2250+6328 (see Fig.~\ref{f:Sp4} in Appendix~\ref{s:Sample_spectra} for an example). 
Their hydrogen and helium emission lines are clearly visible. Interestingly, the algorithm classified both CVs as DQs. We can only speculate that the algorithm somehow confused the pattern of emission lines superimposed on the thermal continuum with C$_{2}$ Swan bands.

\section{General analysis}
\label{s:Gen_ana}

In this section, we present a general analysis of the spectra of the 255 objects observed by the GTC. We first ascertain the capabilities of our Random Forest algorithm and the correctness of its predictions. Subsequently, we unveil the true nature of the `massive DB' subpopulation described in both \cite{Garcia2025} and \cite{Vincent2024} and its properties. Finally, we analyse the location of our identified objects in the \G HR diagram, compare the characteristics of our identified warm and hot DQs against the results of previous works, and give first-order estimates of the magnetic fields for selected objects. 

\subsection{Machine learning performance}
\label{ss:Ascert}

One of the aims of our observational campaigns was to ascertain the accuracy of automatic classifications based on machine learning algorithms. In this subsection, we compare the GTC-based spectral types (hereafter observational spectral types) of 143 objects to those provided in \cite{Garcia2025} and \cite{Vincent2024} and derive their classification metrics. We also compare both automatic classifications against each other.

\subsubsection{Comparison with \cite{Garcia2025}}
\label{sss:Comparison_Garcia}
\begin{figure}[ht]
\centering
\includegraphics[width=1\columnwidth,trim=50 0 50 30, clip]{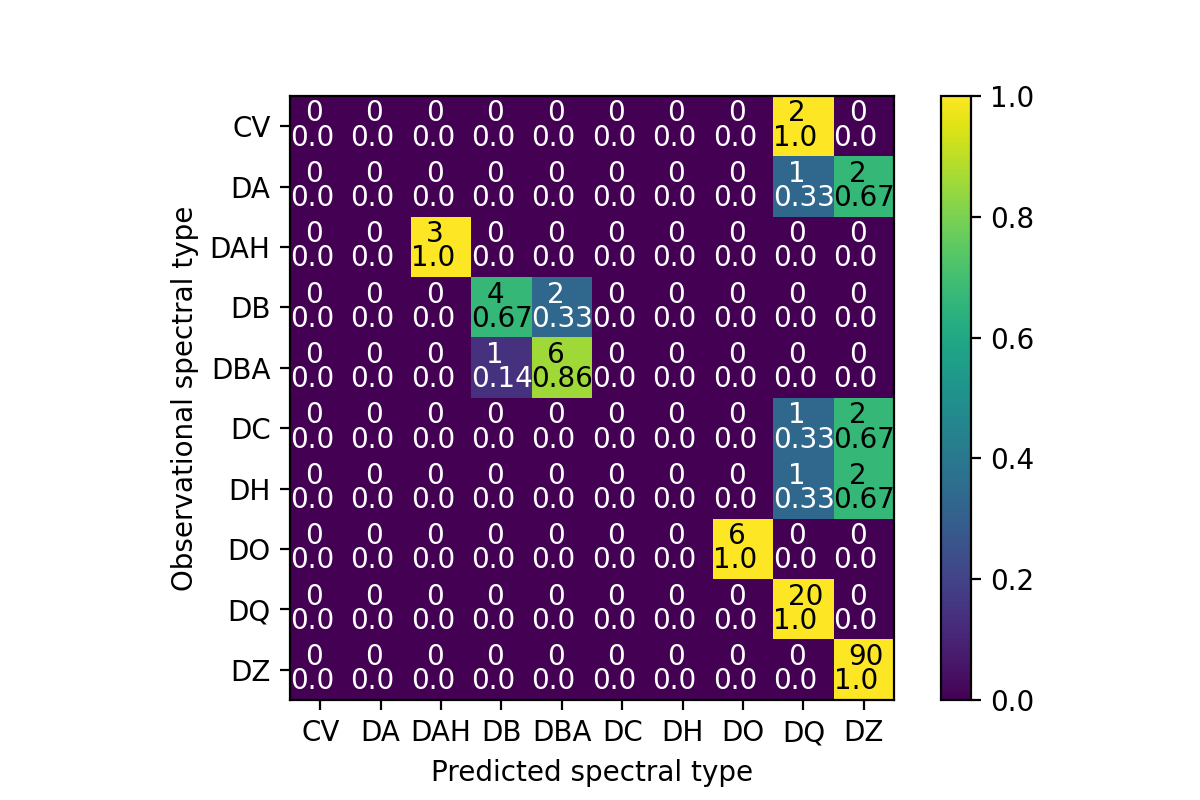}
   \caption{Observed versus predicted \citep{Garcia2025} spectral type confusion matrix for the 143 white dwarfs assigned a spectral type. `DH' designates unclassified magnetic white dwarfs.}
    \label{f:cmatrices}
\end{figure}

A total of 143 white dwarfs that had been previously classified by our Random Forest algorithm as DAH (3), DBs (5), DBAs (8), DOs (6), DQs (25) or DZs (96) were assigned a spectroscopic type based on our optical GTC spectra. 140 of these were assigned an observational spectral type, while the remaining 3 objects are classified as strongly  magnetic with no main spectral type assigned. A confusion matrix comparing the predicted and observational classifications is shown in Fig.~ \ref{f:cmatrices}. Additionally, we present the classification metrics in Table~\ref{tab:Metrics_todo}.

\begin{table}[h!]
    \caption[]{Classification metrics for the observational versus \cite{Garcia2025} comparison.}
    \label{tab:Metrics_todo}
\begin{center}
    \begin{tabular}{lcccccc}
            \noalign{\smallskip}
            \noalign{\smallskip}
        Metric & DAH & DB & DBA & DO & DQ & DZ \\
            \noalign{\smallskip}
            \hline
            \noalign{\smallskip}
        Recall & 1 & 0.67 & 0.86 & 1 & 1 & 1 \\
            \noalign{\smallskip}
            \hline
            \noalign{\smallskip}
        Precision & 1 & 0.8 & 0.75 & 1 & 0.80 & 0.94  \\
            \noalign{\smallskip}
            \hline
            \noalign{\smallskip}
        F1 score & 1 & 0.73 & 0.8 & 1 & 0.89 & 0.97 \\
            \noalign{\smallskip}
            \hline
            \noalign{\smallskip}
    \end{tabular}
\end{center}
\end{table}

As it can be seen from the practically diagonal matrix, the results show a very good agreement between the machine-learning predictions and the  visually assigned spectroscopic classifications, with 129 objects (90.21\%) assigned to the correct spectral type. All observed DAHs, DOs, DQs and DZs had been correctly predicted by our algorithm. The main sources of discrepancies are the identification of DC and DA white dwarfs, as well as misclassification at the DB/DBA subtype level. Additionally, two objects ended up being CVs (Section\,\ref{ss:CVs}), a class not included in the training set and therefore impossible to classify correctly.

\subsubsection{Comparison with \cite{Vincent2024}.}
\label{ss:Comparison_Olivier}

We also compared the observational classification of these 143 objects with the automatic classification by  \cite{Vincent2024}. This classification only took into account the DA, DB, DC, DO, DQ and DZ primary types; no subtypes such as DAH or DBA were considered. Additionally, we assigned the objects with uncertain classification (e.g. DA:) to that primary type.

\begin{figure}[ht]
\centering
\includegraphics[width=1\columnwidth,trim=50 0 50 30, clip]{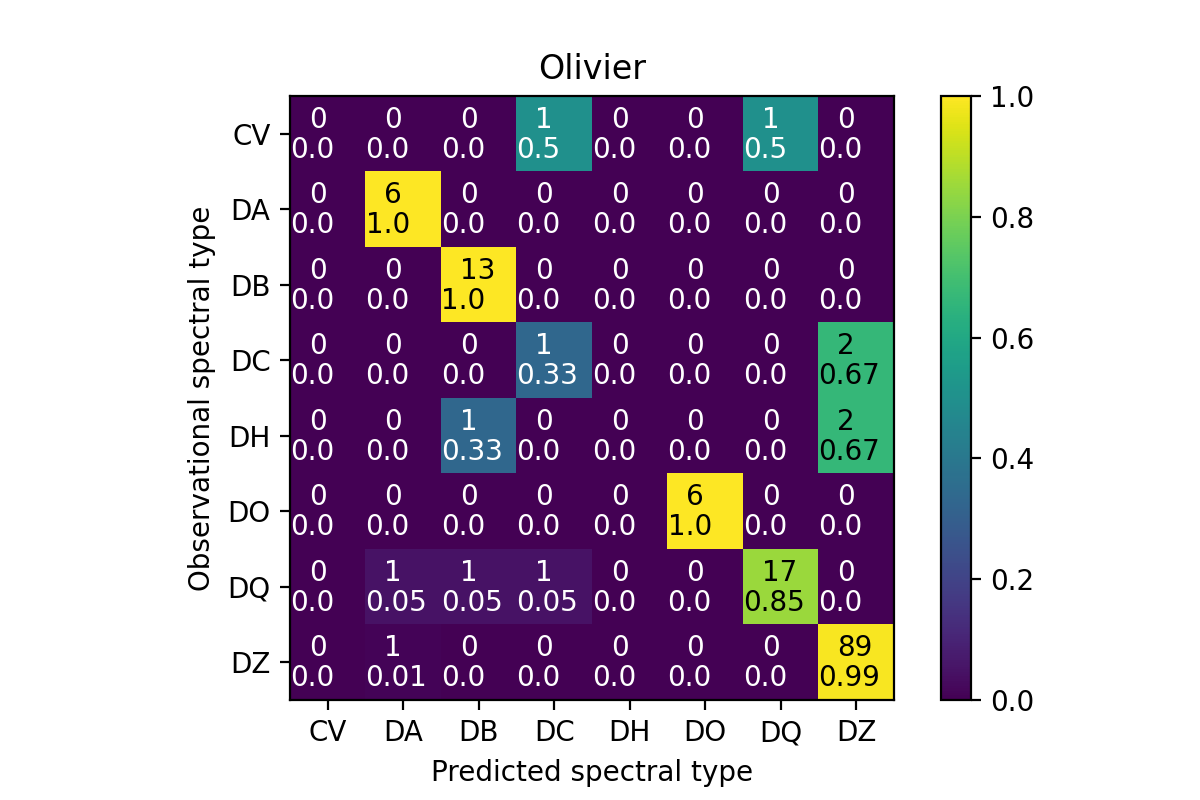}
\caption{Observed versus predicted spectral type \citep{Vincent2024} confusion matrix for the 143 white dwarfs assigned a spectral type. `DH' designates unclassified magnetic white dwarfs.}
    \label{f:cmatrices_Olivier}
\end{figure}

The results are shown in Fig.~\ref{f:cmatrices_Olivier}. A total of 132 out of 143 (92.31\% accuracy) are correctly classified. As above, CVs and magnetic objects could not have been correctly classified due to their absence from the training set. All DAs and one DC are correctly classified as such. However, three DQs and one DZ appear misclassified. Classification metrics for this comparison are shown in Table~\ref{tab:Metrics_Olivier}.

\begin{table}[h!]
    \caption[]{Classification metrics for the observational versus \cite{Vincent2024} comparison.}
    \label{tab:Metrics_Olivier}
\begin{center}
    \begin{tabular}{lcccccc}
            \noalign{\smallskip}
            \noalign{\smallskip}
        Metric & DA & DB & DBC & DO & DQ & DZ \\
            \noalign{\smallskip}
            \hline
            \noalign{\smallskip}
        Recall & 1 & 1 & 0.33 & 1 & 0.85 & 0.99 \\
            \noalign{\smallskip}
            \hline
            \noalign{\smallskip}
        Precision & 0.75 & 0.87 & 0.33 & 1 & 0.94 & 0.96  \\
            \noalign{\smallskip}
            \hline
            \noalign{\smallskip}
        F1 score & 0.86 & 0.93 & 0.33 & 1 & 0.89 & 0.97 \\
            \noalign{\smallskip}
            \hline
            \noalign{\smallskip}
    \end{tabular}
\end{center}
\end{table}

\subsubsection{\cite{Garcia2025} versus \cite{Vincent2024} classifications comparison}
\label{sss:Oli_vs_me}

While at first glance the classification by \citet{Vincent2024} appears to be slightly more accurate than that by \citet{Garcia2025}, it is important to emphasise that \citet{Vincent2024} do not perform spectral subtype classifications. Thus, if DAHs are grouped with DAs and DBAs with DBs, both works correctly classify 132 of the 143 objects (92.31\%). This demonstrates the robustness of independent machine-learning algorithms.

\subsection{Massive DBs' true nature}
\label{ss:True}

For the 2024B, 2025A and 2025B semesters, 170 objects belonging to the `massive DB' sample drawn from the \cite{Garcia2025} and \cite{Vincent2024} catalogues were selected for follow-up spectroscopy. Of these, 112 (65.88\%) were eventually observed and assigned a spectral type, a sample that is statistically significant to derive meaningful conclusions.

Prior to the observational campaigns, we conducted a preliminary search for these objects in the MWDD. Only a few objects were found; at most 20\% were registered as any DB subtype, with warm DQs and magnetic white dwarfs constituting the majority. In our observations, meanwhile, we identify only five DBs (4.46\%) among the 112 classified objects. In contrast, we find 1 DAQ, 23 warm DQs, 4 hot DQs, and 51 magnetic objects. Of these, 11 are DAHs, 4 are hot DQHs, and 36 remain unclassified. A detailed breakdown of these results is provided in Table~\ref{tab:MDB_breakdown}.

\begin{table}[h!]
    \caption[]{Breakdown by spectral type of the `massive DB'  white dwarfs observed in the different spectroscopic campaigns.}
    \label{tab:MDB_breakdown}
\begin{center}
    \begin{tabular}{lcc}
            \noalign{\smallskip}
            \noalign{\smallskip}
        Spectral type (visual) & Number & Percentage\\
            \noalign{\smallskip}
            \hline
            \noalign{\smallskip}
        DA & 6 & 5.36\% \\
            \noalign{\smallskip}
            \hline
            \noalign{\smallskip}
        DAB & 2 & 1.79\% \\
            \noalign{\smallskip}
            \hline
            \noalign{\smallskip}
        DAH & 11 & 9.82\% \\
            \noalign{\smallskip}
            \hline
            \noalign{\smallskip}
        DAQ & 1 & 0.89\% \\
            \noalign{\smallskip}
            \hline
            \noalign{\smallskip}
        DB & 3 & 2.68\% \\
            \noalign{\smallskip}
            \hline
            \noalign{\smallskip}
        DBA & 2 & 1.79\% \\
            \noalign{\smallskip}
            \hline
            \noalign{\smallskip}
        DC & 24 & 21.43\% \\
            \noalign{\smallskip}
            \hline
            \noalign{\smallskip}
        warm DQ & 8 & 7.14\% \\
            \noalign{\smallskip}
            \hline
            \noalign{\smallskip}
        warm DQA & 14 & 12.50\% \\
            \noalign{\smallskip}
            \hline
            \noalign{\smallskip}
        warm DQZA & 1 & 0.89\% \\
            \noalign{\smallskip}
            \hline
            \noalign{\smallskip}
        hot DQ & 4 &3.57\% \\
            \noalign{\smallskip}
            \hline
            \noalign{\smallskip}
        Unclassified magnetic & 36 & 32.14\%  \\
            \noalign{\smallskip}
            \hline
            \hline
            \noalign{\smallskip}
        Total magnetic & 51 & 45.54\%  \\
            \noalign{\smallskip}
            \hline
            \hline
            \hline
            \noalign{\smallskip}
       Total & 112 & 100\% \\
    \end{tabular}
\end{center}
\end{table}

These proportions indicate that, despite its designation, the `massive DB' subpopulation is not composed of genuinely massive DB white dwarfs, but rather of a collection of objects with peculiar spectra that are misclassified as DBs. Neither these very rare spectral types nor the strongly magnetic objects were included in the training sets, and therefore these classifications were not available for the algorithm. As a result, these objects were classified as the closest available category, which happened to be DBs.

\subsection{DA analysis}
\label{ss:DA_analysis}

The location in the HR diagram of the observed DAs separates them into two groups (see Fig.~\ref{f:HRs_analysis} in Appendix~\ref{s:HR_diagrams}, top left panel): the massive and the non-massive samples. The non-massive DAs are found along the upper part of the white dwarf locus, and are relatively cool objects. On the other hand, most of the massive DAs are located above the Q branch, with a few within it, and one DAB and one DAH below it.

Additionally, 4 DAs are massive and very hot ($G_{\rm BP}-G_{\rm RP}<-0.36$). These four objects may provide examples of the hottest end of the DA$\rightarrow$DAQ$\rightarrow$DQA evolutionary sequence for warm DQs proposed by \cite{Kilic2024}. For further details, see Sect.~\ref{sss:DQ_HR}.

\subsection{DB analysis}
\label{ss:DBs_analysis}

As it can be seen in the {\em Gaia} HR diagram (Appendix~\ref{s:HR_diagrams}, Fig.~\ref{f:HRs_analysis}, top right panel), most of the identified DBs occupy a narrow range in colour around $G_{\rm BP} - G_{\rm RP} \approx 0$. Five of them (three DBs and two DBAs) are massive DB white dwarfs. Although they do not form a statistically meaningful sample, these objects still provide valuable information on massive DB white dwarfs. For instance, if they pulsate, their interiors can be studied through asteroseismology. Pulsating DB white dwarfs are expected to be hotter and considerably less crystallised than pulsating DAs \citep{Corsico2021}, thus allowing us to probe their deep interiors.

Interestingly, four of the objects have unusually high luminosities and are hence located slightly above the main cooling track. Currently, we have no observations that allow us to elucidate whether they are low-mass white dwarfs or unresolved binary objects.

\subsection{DC analysis}
\label{ss:DC_analysis}
As stated in Sect.~\ref{ss: DCs}, 24 of the DCs in our sample are hot objects with $G_{\rm BP} - G_{\rm RP} < -0.1$ (see Fig.~\ref{f:HRs_analysis}, Appendix~\ref{s:HR_diagrams}, second row, left panel). However, these two facts are contradictory, since such high temperatures should be sufficient for these objects to display spectral lines. The most likely explanation is that these objects are highly magnetic, with magnetic field strengths high enough to suppress the spectral lines in the optical range \citep{Jewett2024}. However, no polarimetric observations were carried out during our campaigns that could measure their magnetic fields, and therefore we have no direct measurements of their field strengths.

\subsection{DO analysis}
\label{ss:DO_analysis}
We have identified two pure DOs and four DOZs. DOZ objects are a short intermediate phase between pre-degenerate PG 1159 objects and pure DOs, which takes place during the gravitational settling of heavier elements \citep{Bedard2022b}. Therefore, DOZ white dwarfs would be expected to be hotter than pure DOs, which is not what we observe in the {\em Gaia} HR diagram (Fig.~\ref{f:HRs_analysis} in Appendix~\ref{s:HR_diagrams}, top right panel). Although six objects do not constitute a statistically significant sample, it is worth noting that \citet{Unglaub2000} suggest that different initial compositions of PG~1159 objects might cause the PG~1159$\rightarrow$DO transition to occur at different temperatures. By extension, the transitional DOZ phase would also take place at different temperatures, hence colours, for different objects. This difference would explain the obtained distribution.

\subsection{Carbon rich stars}
\label{ss:DQ_analysis}

Warm and hot DQs are rare and poorly understood objects. In the last years, many works \citep[e.g.][]{Koester2019,Coutu2019,Kilic2024,Kilic2025} have shed light on the properties, parameters and evolution of these white dwarfs.

In this section, we analyse the different DQ subpopulations we have identified. This includes both their location in the \G HR diagram and their kinematics, and a comparison to the results of recent studies such as \cite{Coutu2019} and \cite{Kilic2025}.

\subsubsection{The DQ subpopulations}
\label{sss:DQ_subpops}

White dwarfs with carbon features do not constitute a homogeneous subpopulation. Rather, they separate into two distinct subpopulations: on the one hand, the cool, classical DQs; on the other hand, the warm and hot DQs  (see, for instance,  \citealt{Dufour2013,Fortier2015,Coutu2019,Koester2019}).

The distinction between the cool DQ and the warm/hot DQ subpopulations is not based only on their different temperatures, but also on their stellar parameters (e.g. mass), atmospheric composition and proposed origins. Both warm and hot DQs are more massive than classical, cool DQs (see, e.g. \citealt{Koester2019,Coutu2019,Kilic2025,OuldRouis2026}). A very high proportion of hot DQs is also found to be magnetic \citep{Dufour2013}; magnetic warm DQs, however, are far more rare \citep{OuldRouis2026}; see \cite{Williams2013} for an example. As for their atmospheres, warm and hot DQs are dominated by carbon \citep{Dufour2007, Dufour2008} with hydrogen as a trace element \citep{Koester2019,Kilic2025}; whereas cool DQ atmospheres are helium dominated \citep{Bues1973}, and carbon appears in their spectra as a result of convective dredge-up \citep{Koester1982, Pelletier1986}. Finally, the kinematics of warm DQs, more specifically their high velocity dispersion, reveal an incoherence between their cooling and kinematic ages. That is, warm DQs cooling ages are small compared with the kinematic ages derived from their motions \citep{Dunlap2015,Coutu2019,Kawka2023}.

This combination of higher-than-average masses, peculiar atmospheric composition, a high incidence of magnetic fields, and unusual kinematics suggests that a merger-remnant origin may represent one of the most plausible formation channels for warm and hot DQs \citep{Dunlap2015,Coutu2019,Kawka2023}. Cool DQs, on the other hand, are the result of single object evolution and are one of the final proposed steps in the evolutionary sequence of helium-rich objects (\cite{Bedard2022}, see also Fig.~13 in \citealt{Bedard2024}). In the following subsections, we analyse these aspects for the warm and hot DQs in our sample, focusing on whether our results agree with previous studies.

\subsubsection{\G HR diagram location}
\label{sss:DQ_HR}

Once the DQs are placed in the {\em Gaia} HR diagram (Appendix~\ref{s:HR_diagrams}, Fig.~\ref{f:HRs_analysis}; second row, right panel: all DQs; third row, left panel: warm and hot DQs), the two subpopulations occupy markedly different regions. Cool DQs spread along the main cooling track, with $G_{\rm BP}-G_{\rm RP}$ colours larger than $\simeq$0.5 mag. Warm DQs are concentrated on the Q branch and the crystallization sequence. Finally, hot DQs are located in the bluer end of the white dwarf locus ($G_{\rm BP}-G_{\rm RP}\leq-0.3$) and approximately one magnitude above the Q branch.

The location of our warm DQs is coincident and compatible with those analysed by \citealt{Kilic2025} (see top panel of their Fig.~11, as well as their Fig.~12). That is, warm DQs appear along the Q branch and crystallization sequences, while hot DQs have not reached it yet. Our single identified DQZA also lies on the Q branch; the DAQ we identified lies there as well, but at a bluer location.

Lastly, we highlight the cases of warm DQs J0517+2633, J0540+3205, J1902+2607, J2024-0234, J2046+3830, and J2323-0623, since, in addition to \ion{C}{I} spectral lines, they also display shallow Swan bands centered around $\lambda\lambda\lambda=4370,\,4700,\,5100\,\AA$ (an example spectrum can be consulted in Appendix~\ref{s:Sample_spectra}, Fig.~\ref{f:Sp3}). These objects, located at the cooler end of the Q branch and displaying warm and cool DQ characteristics at the same time, fall under the warm DQ classification according to the definition in \cite{Koester2019}.

In \cite{Kilic2024}, two evolutionary channels are proposed for warm and hot DQs. In the first one, DA$\rightarrow$DAQ$\rightarrow$warm DQA, a thin hydrogen atmosphere would get progressively diluted by the carbon-rich envelope as the white dwarf cools. In the second, hot DQ(A)$\rightarrow$warm DQ(A), a hot DQ would progressively cool into a warm DQ. We believe the six objects described in the previous paragraph may provide examples of the final stages of these evolutionary channels.

\subsubsection{Magnetic incidence in warm and hot DQs}
\label{sss:Magnetism}

A very high proportion of hot DQs are reported as magnetic. For instance, \cite{Dufour2013} found 10 magnetic objects in a sample of 14 hot DQs (71.43\%). On the other hand, magnetic fields in warm DQs are more strange, with only a few known examples. Different processes, such as white dwarf-subgiant mergers or magnetic field dissipation, have been invoked as possible explanations \citep{OuldRouis2026}.

In our classification, we find that all four hot DQs (100\%) harbour magnetic fields. Their intensities, however, are varied. For instance, the splittings in J0034+7130 is of only a few $\AA$, J0420+6450 has a wavelength difference between the split components of a few tens of $\AA$; while band-like structures appear in the spectra of J0446+7227 and J0642+0632. An estimate for the magnetic field of J0420+6450 can be found in Sect.~\ref{sss:DQH_analysis}.

In contrast to this result, no magnetic warm DQs have been found among our 29 identifications. The incidence of magnetism in hot and warm DQs, therefore, is both compatible and in agreement with previous results.

\subsubsection{Kinematics of the DQ population}
\label{sss:Kinematics}

Recent studies suggest that the warm and hot DQ subpopulations display unusual kinematics. For instance, \citet{Dunlap2015} and \citet{Kilic2025} report inconsistencies between their kinematic ages and cooling ages, with the former being much greater than the latter. In this sense, a high proportion of the warm DQ population displays unusually high tangential velocities ($v_{\rm tan}\geq50$ km s$^{-1}$, used in \cite{Wegg2012} as a kinematic `smoking gun' signature of merger origin for high-mass white dwarfs). \cite{Coutu2019} finds that 10 out of 22 warm DQs (45.5\%) in their sample showed $v_{\rm tan}\geq50$ km s$^{-1}$. \cite{Kilic2025} reports 41 out of 75 warm DQs (54.7\%) above this value.

We obtained the tangential velocities of all white dwarfs in our observational campaigns and show them in a colour–$v_{\rm tan}$ diagram (Fig.~\ref{f:Kinematics}), following Fig.~13 of \citealt{Kilic2025}.  We find that 52 out of the 322 objects in our complete observational sample have $v_{\rm tan}\geq\,50$ km s$^{-1}$. Of them, 13 are warm DQs. Taking into account that we have found a total of 33 warm and hot DQs, of which 11 are pure DQs, 16 are DQAs, 1 is a DQZA, 1 is a DAQ, and 4 are hot DQs, this shows that 39.4\% of our warm DQ sample has $v_{\rm tan}\geq\,50$ km s$^{-1}$. Although the proportion of warm DQs with unusual kinematics is not as high as the one found in \cite{Coutu2019} or \cite{Kilic2025}, it is nevertheless unusually high. 

For cool DQs, only three out of 15 (20\%), two of which are DQpecs, show $v_{\rm tan}\geq\,50$ km s$^{-1}$, a fraction close to the one found for the whole sample (52 out of 322 objects, 16.15\%). The cool DQ kinematics are, therefore, compatible with single object evolution.

\begin{figure}[ht]
\centering
\includegraphics[width=1\columnwidth,trim=-20 0 0 0, clip]{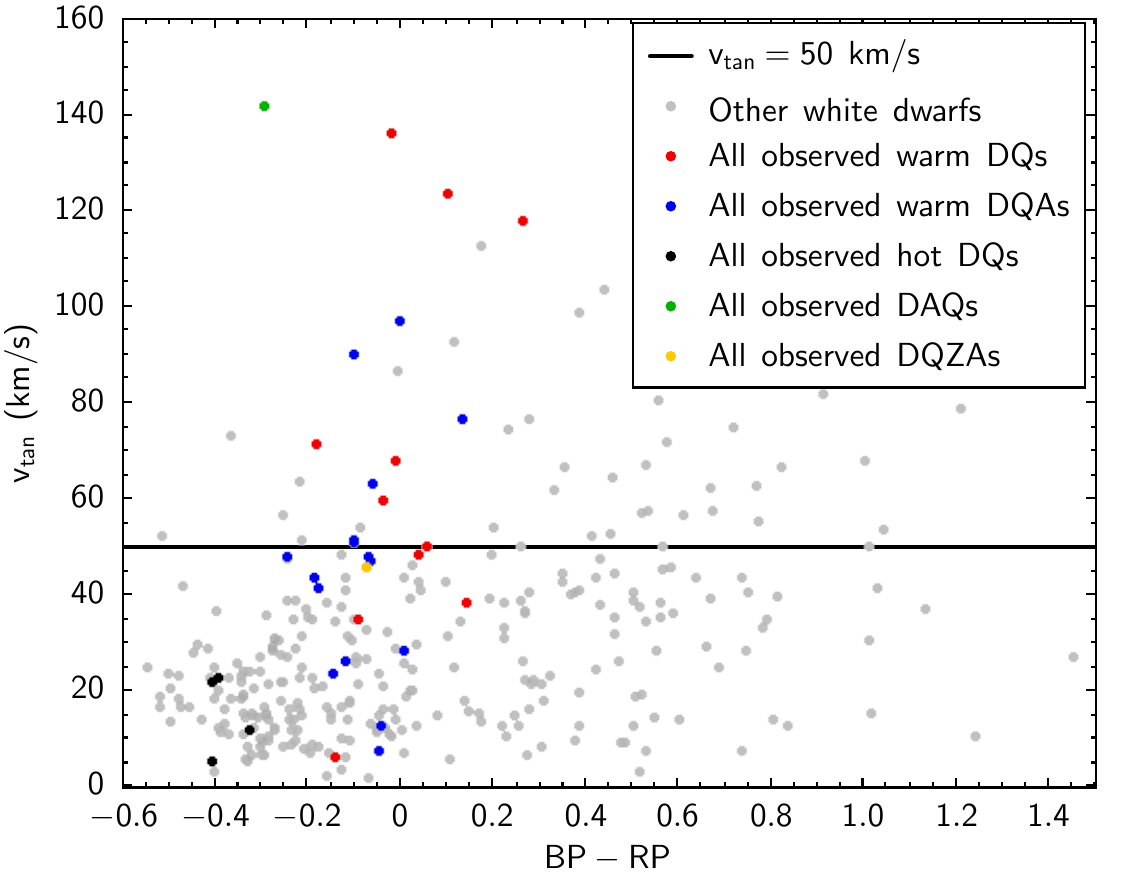}
\caption{Colour-tangential velocity diagram for our warm DQs. It can be seen that thirteen of our warm DQs are above the $v_{\rm tan}\geq\,50$ km s$^{-1}$ limit.}
    \label{f:Kinematics}
\end{figure}

\subsection{DZ analysis}
\label{ss:DZ_analysis}

The 90 identified DZs are displayed in the \G HR diagram in Appendix~\ref{s:HR_diagrams}, Fig.~\ref{f:HRs_analysis}, third row, right panel. They are found mainly along the B branch. As discussed above, we identify the \ion{Ca}{II} H and K lines in all objects, while other metals such as Mg, Fe or Na are more infrequent.

 Ten of our objects display weak hydrogen signatures as well as metallic lines, which is thought to have been accreted from water-rich bodies \citep{Jura2012,GentileFusillo2017}. 
 
DZ white dwarfs with helium features are exceedingly rare objects, with only four cases (two DZAB, one DZBAH, and one DZO) listed in the MWDD. In our campaigns, we have identified three more such objects, namely two DZBAs (J0437+0051 and J0745+5551) and 1 DZAB (J0452-0214), therefore increasing the amount of known objects by 75\%. When represented in the \G HR diagram (third row, right  of Fig.~\ref{f:HRs_analysis} in Appendix~\ref{s:HR_diagrams}), these objects are hot ($G_{\rm BP}-G_{\rm RP}\sim0$), as expected from the presence of \ion{He}{I} lines.

\subsection{Magnetic object analysis}
\label{ss:DXH_analysis}

In addition to the spectroscopically confirmed magnetic white dwarfs discussed above, our sample also contains a number of objects displaying clear magnetic signatures. Of the unclassified magnetic objects, 36 are hot and massive, while three are cooler and lie along the main cooling track. When represented in the {\em Gaia} HR diagram, these objects are located mostly above, albeit close to, the hotter end of the Q branch, with only a minority lying directly on it. This behaviour is consistent with the location of magnetic white dwarfs registered in the MWDD. However, in the absence of polarimetric observations and without a reliable spectral classification, the magnetic field strengths of these objects cannot be determined from spectroscopy alone.

The {\em Gaia} HR diagram location of these objects can be consulted in the bottom left panel of Fig.~\ref{f:HRs_analysis} in Appendix~\ref{s:HR_diagrams}, while the bottom right panel shows a zoom of the Q branch region, highlighting the DAHs, hot DQHs, and the massive unclassified magnetic objects.

In the following subsections we estimate the magnetic field strengths of the spectroscopically classified magnetic white dwarfs in our sample, namely the DAHs, DZHs, and hot DQHs.

\subsubsection{DAH magnetic field estimates}
\label{sss:DAH_analysis}

\begin{figure}[h!]
\centering
    \includegraphics[width=1.0\columnwidth,trim=0 0 30 0, clip]{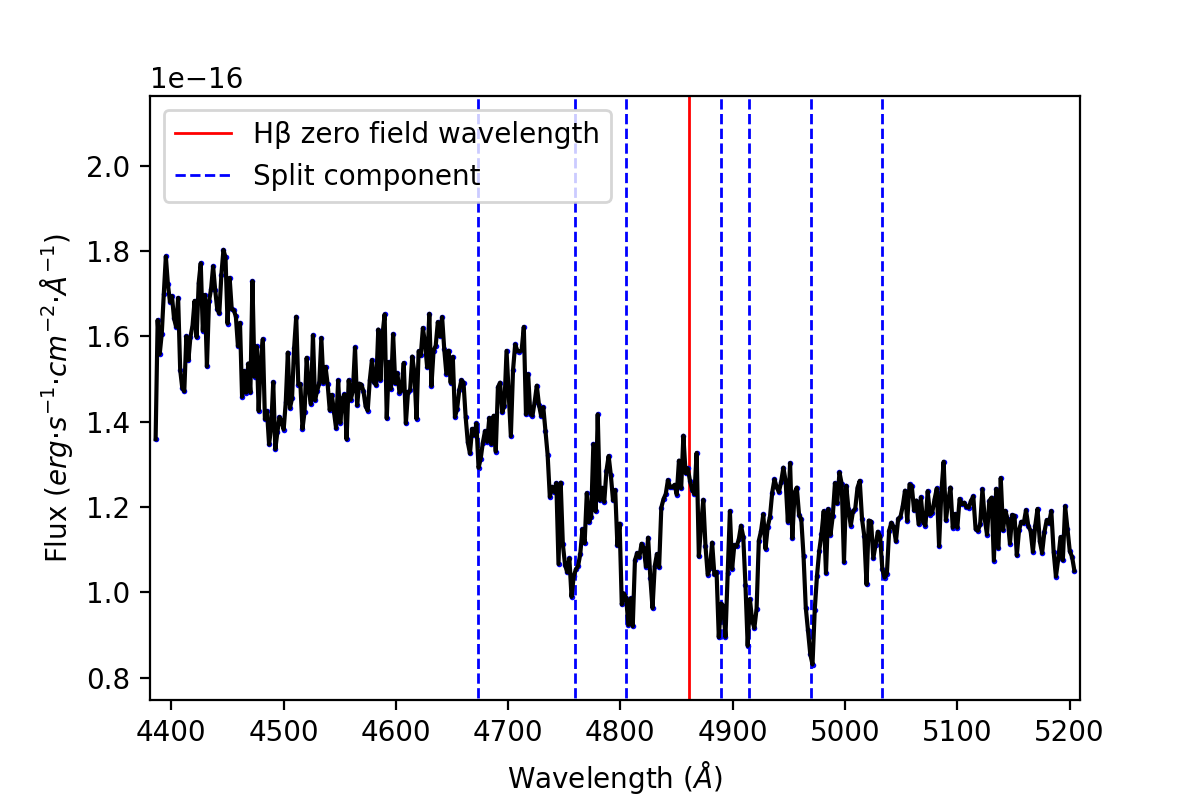}
    \vspace{0.0cm}
    \caption{H$\beta$ Zeeman splitting in DAH J1927+6938, where H$\beta$ splits into seven components.}
    \label{f:Se7en}
        \vspace{0.0cm}
\end{figure}

Based the magnitude of the Zeeman splitting, we find two different cases, corresponding to different magnetic-field strengths. Three objects display three components in the split H$\beta$ line, whereas the 11 remaining ones show seven. Additionally, the Zeeman splitting of H$\alpha$ is larger for these objects. This indicates a higher magnetic field. An example of the seven-component splitting of the H$\beta$ line is displayed in Fig.~\ref{f:Se7en}.

We also performed a first-order estimate of the magnetic field strengths by fitting Gaussian profiles to the visible components of the H$\alpha$ and H$\beta$ lines. The line center was considered to lie at the mean of the Gaussian fits, and the spectral displacement was computed as the wavelength difference between the lateral and central peaks for both components.

Once we have calculated the spectral displacements, we use Eq.~1 of \cite{Reid2001} to obtain a magnetic field estimate for each displacement, and average them to obtain a final magnetic field for each spectral line. Then, the final magnetic field of the white dwarf is obtained by taking the average of the fields for H$\alpha$ and H$\beta$. In cases where the H$\beta$ line is further divided into seven components, rather than three, only the three components of the H$\alpha$ line are used for magnetic field estimation.

The magnetic fields for each line and the final magnetic field for each of the white dwarfs are presented in Table~\ref{tab:BField}. As expected, the white dwarfs with larger Zeeman splitting have a stronger magnetic field, of the order of 20\,MG, whereas the objects with narrower splitting have magnetic fields of the order of 10\,MG.

\begin{table}[h!]
    \caption[]{Magnetic field estimations for the spectroscopically confirmed DAH white dwarfs.}
    \label{tab:BField}
\begin{center}
    \begin{tabular}{lccc}
            \noalign{\smallskip}
            \noalign{\smallskip}
        Name & $B_{H\alpha}$ (MG) & $B_{H\beta}$ (MG) & $B_{\rm WD}$ (MG) \\
            \noalign{\smallskip}
            \hline
            \noalign{\smallskip}
        J0318+7923 & 9.8$\pm$0.7 & 11.0$\pm$0.6 & 10.4$\pm$0.7  \\
            \noalign{\smallskip}
            \hline
            \noalign{\smallskip}
        J0424+1054 & 9.28$\pm$0.18 & 8.49$\pm$0.17 & 8.89$\pm$0.17  \\
            \noalign{\smallskip}
            \hline
            \noalign{\smallskip}
        J0547-1250 & 10.66$\pm$0.16 & 10.6$\pm$0.2 & 10.6$\pm$0.2 \\
            \noalign{\smallskip}
            \hline
            \hline
            \noalign{\smallskip}
        J0043-1000 & 18.7$\pm$0.4 & - & 18.7$\pm$0.4 \\
            \noalign{\smallskip}
            \hline
            \noalign{\smallskip}
        J0528+5955 & 22.2$\pm$0.6 & - & 22.2$\pm$0.6 \\
            \noalign{\smallskip}
            \hline
            \noalign{\smallskip}
        J0758+3544 & 13.6$\pm$0.9 & - & 13.6$\pm$0.9 \\
            \noalign{\smallskip}
            \hline
            \noalign{\smallskip}
        J0911-1107 & 23.3$\pm$0.9 & - & 23.3$\pm$0.9 \\
            \noalign{\smallskip}
            \hline
            \noalign{\smallskip}
        J1154+0117 & 18.2$\pm$0.4 & - & 18.2$\pm$0.4 \\
            \noalign{\smallskip}
            \hline
            \noalign{\smallskip}
        J1339-0713 & 19.2$\pm$0.3 & - & 19.2$\pm$0.3 \\
            \noalign{\smallskip}
            \hline
            \noalign{\smallskip}
        J1516+3933 & 16.1$\pm$0.9 & - & 16.1$\pm$0.9 \\
            \noalign{\smallskip}
            \hline
            \noalign{\smallskip}
        J1808+6523 & 23.5$\pm$0.9 & - & 23.5$\pm$0.9 \\
            \noalign{\smallskip}
            \hline
            \noalign{\smallskip}
        J1813+1549 & 20.2$\pm$0.4 & - & 20.2$\pm$0.4 \\
            \noalign{\smallskip}
            \hline
            \noalign{\smallskip}
        J1927+6938 & 19.6$\pm$0.3 & - & 19.6$\pm$0.3 \\
            \noalign{\smallskip}
            \hline
            \noalign{\smallskip}
        J2343+5728 & 19.3$\pm$0.5 & - & 19.3$\pm$0.5\\
            \noalign{\smallskip}
            \hline
            \noalign{\smallskip}
\end{tabular}
\end{center}
\end{table}

Finally, it is worth to note that all of our identified DAHs are hot objects ($G_{\rm BP}-G_{\rm RP}\leq0.02$). This contrasts with the distribution of known DAH objects registered in the MWDD, which span a wide range of temperatures, from the hotter to the colder end of the white dwarf locus. This result is due to our own selection bias, as we mostly observed hot ($G_{\rm BP}-G_{\rm RP}\leq0$) objects, therefore precluding us from observing cooler DAHs.

\subsubsection{DZH magnetic field estimates}
\label{sss:DZH_analysis}

 Following the approach we adopted for the magnetic splitting of the Balmer lines in the previous section, we estimated the magnetic field strengths of this subsample using the splitting of the \ion{Na}{I} doublet by fitting the three peaks to Gaussian profiles and using Eq. 1 in \cite{Reid2001}. Our estimates show four objects with magnetic fields of a few MG. These values can be consulted in Table~\ref{tab:BField_DZH}.

\begin{table}[ht]
    \caption[]{Magnetic field strengths of the identified DZH white dwarfs.}
    \label{tab:BField_DZH}
\begin{center}
    \begin{tabular}{lc}
            \noalign{\smallskip}
            \noalign{\smallskip}
        Name & $B_{\rm WD}$ (MG) \\
            \noalign{\smallskip}
            \hline
            \noalign{\smallskip}
        J0342+2934 & 2.35$\pm$0.06 \\
            \noalign{\smallskip}
            \hline
            \noalign{\smallskip}
        J0610-1402 & 1.30$\pm$0.14 \\
            \noalign{\smallskip}
            \hline
            \noalign{\smallskip}
        J0738+7424 & 3.07$\pm$0.11 \\
            \noalign{\smallskip}
            \hline
            \noalign{\smallskip}
        J2357+2747 & 1.45$\pm$0.05 \\
            \noalign{\smallskip}
            \hline
            \noalign{\smallskip}
    \end{tabular}
\end{center}
\end{table}

It is important to note that, in the cases of J0829-0818 and J1543-0247, we could not perform an estimation of their magnetic fields. For J1543-0247, the low SNR of the spectra makes us unable to perform a fit to the split components. In the case of J0829-0818, there are simply so many split components that we cannot reliably identify which lines belong to which element. However, due to the similarity of this spectrum to the one described in \cite{Hollands2023}, and which corresponded to a DZH with a magnetic field of approximately 30\,MG, we believe that the magnetic field of this object could be of the same order. This would make J0829-0818 one of the DZs with the most intense magnetic fields registered.

\subsubsection{Hot DQH magnetic fields estimate}
\label{sss:DQH_analysis}

All four identified hot DQHs are, as expected, very hot objects with colour $G_{\rm BP} - G_{\rm RP} < -0.32$, and are located about one magnitude above the Q branch. We also attempted to estimate the magnetic field strengths of the identified DQHs. However, this was only possible for J0420+6450, for which the field was derived by fitting the split components of the \ion{C}{ii} $\lambda\lambda = 6579, 6582\,\AA$ doublet. As in the case of the \ion{Na}{i} doublet described in Sect.~\ref{sss:DZH_analysis}, this \ion{C}{ii} doublet becomes a triplet in the presence of a strong magnetic field. The resulting estimate for this object is $B = 1.87 \pm 0.11$\,MG.

In the case of J0034+7130, the Zeeman splitting is too weak to reliably identify this split triplet. However, judging by its similarity to the spectra shown in Fig.~2 in \cite{Dufour2013}, and which have magnetic fields of a few hundred kG, we conclude that this object likely harbours a magnetic field weaker than 1\, MG.

Finally, the band structure of the spectra of J0446+7227 and J0642+0632 also prevents the correct identification of the split components. Their fields, however, must be stronger than the one of J0420+6450.

\section{Conclusions and future perspectives}
\label{s:Conc}

In this work, we obtained medium-resolution optical spectra of 255 white dwarfs with GTC over four observing campaigns to investigate the reliability of machine-learning-based spectral classifications derived from \G data, by comparing visual classifications from the GTC spectra with those predicted by \citet{Garcia2023, Garcia2025}, and to clarify the nature of specific subpopulations identified by these methods, in particular the so-called massive ($M \geq 0.95\,\mathrm{M}_\odot$) DB white dwarfs.

This analysis enables both a direct validation of automated classifications through comparison with spectroscopic observations and a detailed characterisation of the underlying white dwarf population. Our main contributions can be summarised as follows: 
\begin{enumerate}
 \item A quantitative evaluation of the performance of machine learning algorithms for spectral classification within a large, \G-selected sample of white dwarfs, providing a direct comparison between automated classifications based on \G data and those obtained from medium-resolution spectroscopy.
 \item A definitive characterisation of the so-called `massive DB' population, showing that only $\sim$4.5\% of these objects are genuine DB white dwarfs, while the majority are magnetic white dwarfs or warm and hot DQs, thus providing the first systematic spectroscopic confirmation of their true nature.
 \item  An observational characterisation of warm and hot DQ white dwarfs, including their location in the \G HR diagram and their kinematic properties, consistent with previous works \citep{Coutu2019, Kilic2024, Kilic2025}, and supporting a merger-origin scenario.
 \item The identification of a significant population of magnetic white dwarfs (63 objects), many of them massive and located near the Q branch, contributing to our understanding of the incidence of magnetism in this region.
 \end{enumerate}

A more detailed breakdown of the results is given below:
\begin{enumerate}
\item A total of 26 DAs, 18 DBs, 27 DCs, 6 DOs, 47 DQs, 90 DZs, 2 CVs, and 39 unclassified magnetic objects have been identified.
\item We confirm the high precision of the Random Forest algorithm described in \cite{Garcia2023} and \cite{Garcia2025} for spectral type classification. Only 14 objects (9.79\%) deviate from the predicted spectral type, 5 of which belong to types not included in the training set.
\item Only 5 of the 112 massive DB white dwarfs ($M_{\rm WD}\geq0.95\,$M$_{\odot}$) pre-classified by machine learning algorithms are true DBs. Of these, three (J0622+6935, J0903-1311 and J1832+0856) are pure DBs, and two (J0737+5728 and J0847+3222) are DBAs. The remaining 107 have been either assigned a different type such as DAs, DAHs, DCs, or warm and hot DQ(A)s, or regarded as unclassified magnetic. 
\item We have detected secondary spectral characteristics in 63 of the GTC spectra that were undetected in the low resolution \G spectra, allowing for a more precise subclassification of these objects. These objects belong to rare and extremely rare spectral types: 2 DABs, 14 DAHs, 1 DAQ, 8 DBAs, 1 DBAZ, 4 DOZs , 16 warm DQAs, 1 DQZA, 7 DZAs, 1 DZAB, 2 DZBAs, and 6 DZHs.
\item The identified subspectral characteristics increase the amount of such known objects in the MWDD by the following amounts: 2 DABs (1.82\%), 9 DAHs (1.37\%), 7 DBAs (1.51\%), 1 DBAZ (2.6\%), 4 DOZs (10.25\%), 12 warm DQAs (16.22\%), 1 DQZA (50\%), 7 DZAs (7.45\%), 1 DZAB (50\%), 2 DZBAs (200\%) and 4 DZHs (20\%). 
\item We find 24 massive ($M_{\rm WD}\geq0.95\,$M$_{\odot}$) and hot ($0.10\geq G_{\rm BP}-G_{\rm RP}$) DCs, which are expected to be highly magnetic.
\item We find, not considering the 24 DCs from the previous point, 63 newly identified magnetic objects. Of these, 14 are DAHs, six are DZHs, four are hot DQHs, and 39 remain unclassified due to their spectral features not coinciding with any known spectral lines. Of them, 51 (11 DAHs, four hot DQHs, and 36 unclassified) are massive ($M\geq0.95\,$M$_{\odot}$). These 51 objects, together with the 24 massive DCs found, will help us reach a better understanding of the origin and incidence of magnetic fields in massive white dwarfs.
\item We confirm 14 new magnetic DAHs objects, 11 of which  are massive ($M_{\rm WD}\geq0.95\,$M$_{\odot}$). From the field estimate described in Sect.~\ref{sss:DAH_analysis}, 3 of the DAHs have magnetic fields of the order of 10\,MG; while the other 11 of the order of 20\,MG.
\item Six new magnetic DZH white dwarfs have been identified. The first-order estimates for their magnetic fields (see Sect.~\ref{sss:DZH_analysis}) reveals them to lie inside the 1-4\,MG range.
\item Four new magnetic hot DQHs (J0034+7130, J0420+6450, J0446+7227 and J0642+0632) with different field strengths have been identified. The magnetic field of J0420+6450 has been estimated to be of the order of 2\,MG.
\item Two new CVs (J2007+1742 and J2250+6328) have been identified.
\end{enumerate}

Regarding the warm and hot DQs, our main findings can be summarised as follows:

\begin{enumerate}
\item A total of 29 warm DQs and four hot DQs have been identified.
\item Of the 29 warm DQs, 11 are pure DQs, 16 are DQAs, 1 is a DQZA (J0328+5806), and one is a DAQ (J0325+2540).
\item Of the warm DQs, six of them show, in addition to \ion{C}{I} lines, weak Swan bands. These may provide examples of the evolution pipelines for warm DQs described in \cite{Kilic2024}.
\item The location of our 29 warm DQs in the \G HR diagram agrees with the findings of \cite{Kilic2025}: on the Q branch and the CO core crystallization sequence.
\item Our four hot DQs are located, approximately, one magnitude above the Q branch.
\item The magnetic fraction for hot and warm DQs is in agreement with previously published results. 
\item We find 13 of our 33 (39.4\%) warm and hot DQs have unusually high tangential velocities ($v_{\rm tan}\geq50\,$km s$^{-1}$). Our kinematics analysis results also agree with the results of \cite{Coutu2019} and \cite{Kilic2025}.
\item Our only identified DAQ (J0325+2540) is located in the Q branch and shows a very high ($v_{\rm tan}\geq140$ km s$^{-1}$) tangential velocity. This is in agreement with already published DAQs and warm DQs differing only in their atmospheric carbon abundances \citep{Kilic2024}.
\end{enumerate}

Future perspectives on the performed work are two-fold:

\begin{enumerate}
\item While new insights have been gained on massive magnetic objects, as well as warm and hot DQs, more information is still needed in regard to their relation with the rest of the population of the Q branch. For this purpose, new spectroscopical campaigns have been proposed, spanning objects from the Q branch with no spectral type distinction. These data will be complemented with the data releases from the DESI \citep{DESI2019}, 4MOST \citep{4MOST2019}, and WEAVE \citep{WEAVE2022} surveys.

\item Despite some of the magnetic `massive DBs' being identified as DAH and hot DQH white dwarfs, 36 out of 51 (70.59\%) of them still lack a reliable spectral type. Polarimetric observations in order to determine their magnetic field intensity would be of help in this task. 

\end{enumerate}

\begin{acknowledgements}
We acknowledge support from MINECO under the PID2023-148661NB-I00 grant and by the AGAUR/Generalitat de Catalunya grant SGR-386/2021. Enrique Miguel García Zamora also acknowledges financial support from Banco de Santander, under a Becas Santander Investigación/Ajuts de Formació de Professorat Universitari (2022\_FPU‐UPC\_16) grant. Based on observations made with the Gran Telescopio Canarias (programmes GTC20-23B, GTC28-24B, GTC36-25A, GTC8-25B), installed in the Spanish Observatorio del Roque de los Muchachos of the Instituto de Astrofísica de Canarias, in the island of La Palma.

\end{acknowledgements}

\section*{Data Availability Statement}
The data underlying this article are available in the article.  Supplementary material will be shared on reasonable request to the corresponding author. Table 3 is only available in electronic form at the CDS via anonymous ftp to \url{cdsarc.u-strasbg.fr (130.79.128.5)} or via \url{http://cdsweb.u-strasbg.fr/cgi-bin/qcat/J/A+A/}.

\bibliographystyle{aa}
\bibliography{Spectra}


\begin{appendix}

\begin{figure*}[h!]
\section{\G colour-magnitude diagrams for the visually identified spectral types and subtypes}
\label{s:HR_diagrams}

\centering
    \includegraphics[width=0.8\columnwidth,trim=0 0 0 0, clip]{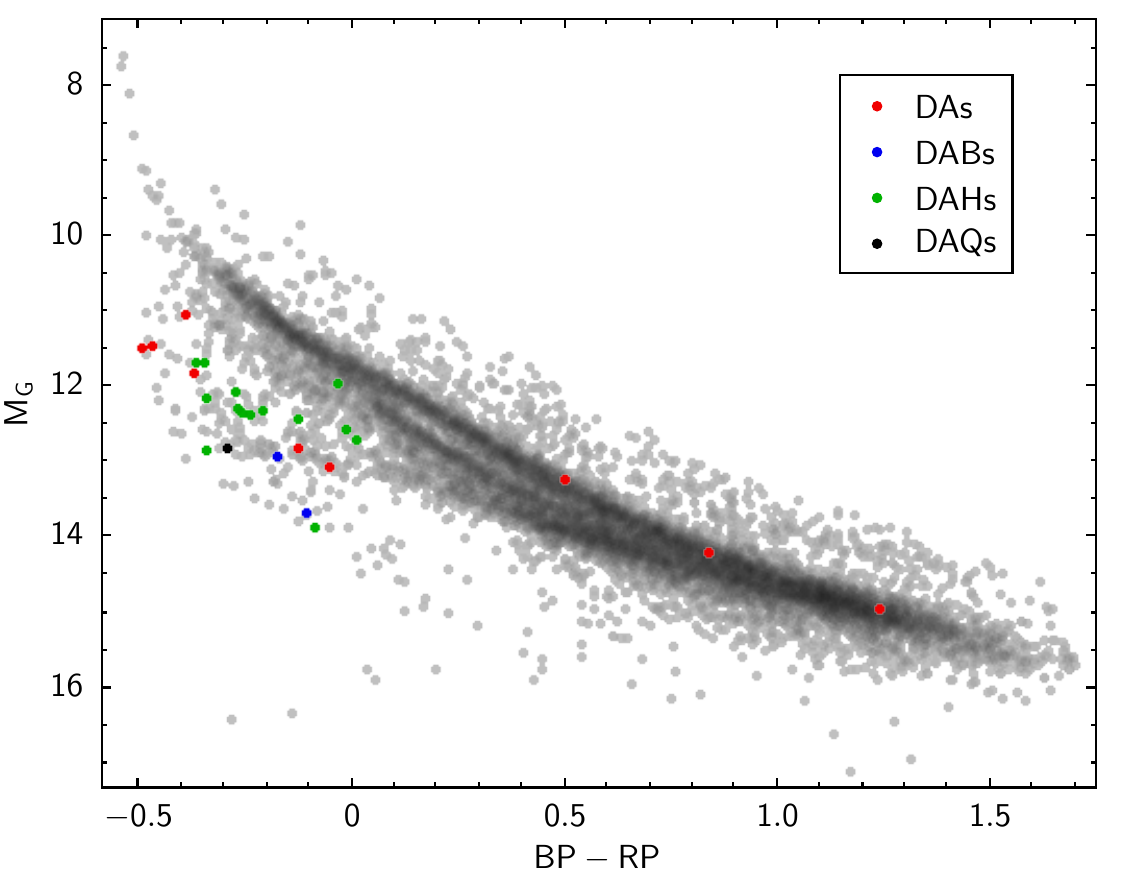}
    \includegraphics[width=0.8\columnwidth,trim=0 0 10 0, clip]{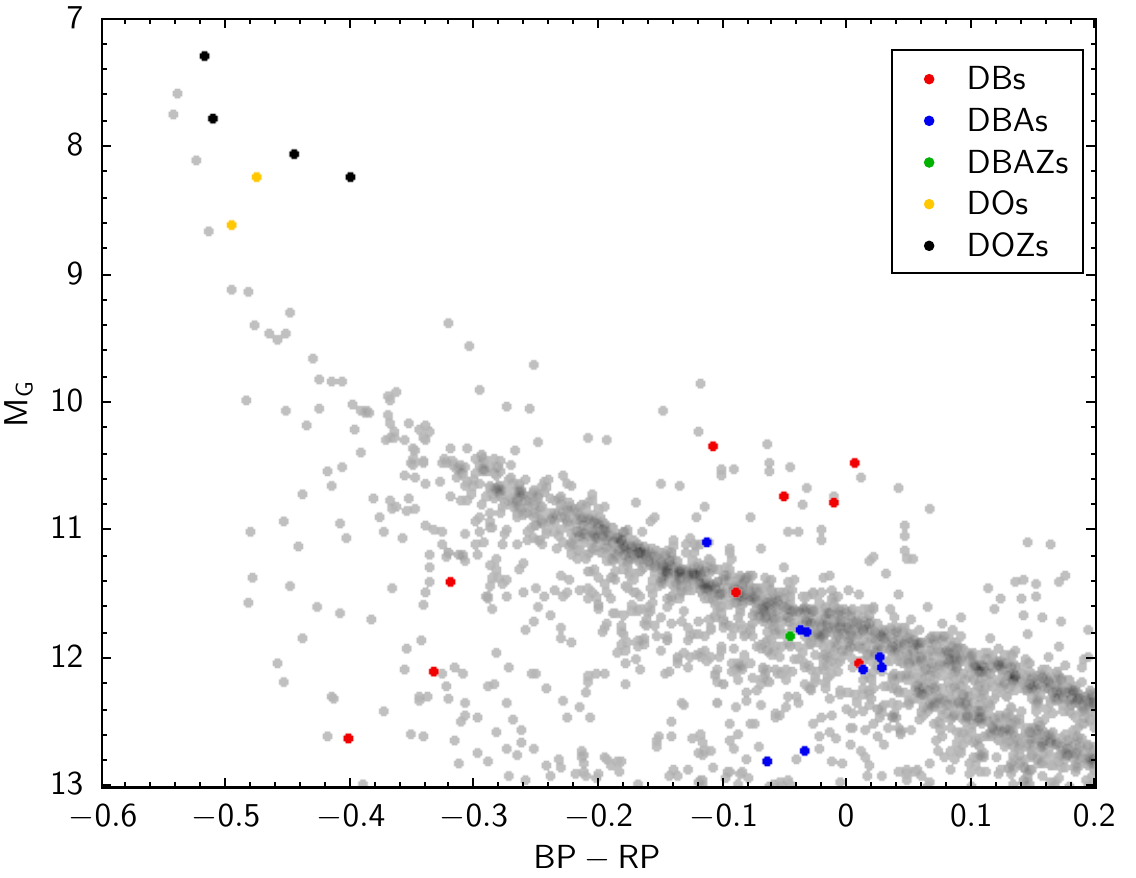}
    \centering
    \includegraphics[width=0.8\columnwidth,trim=0 0 0 0, clip]{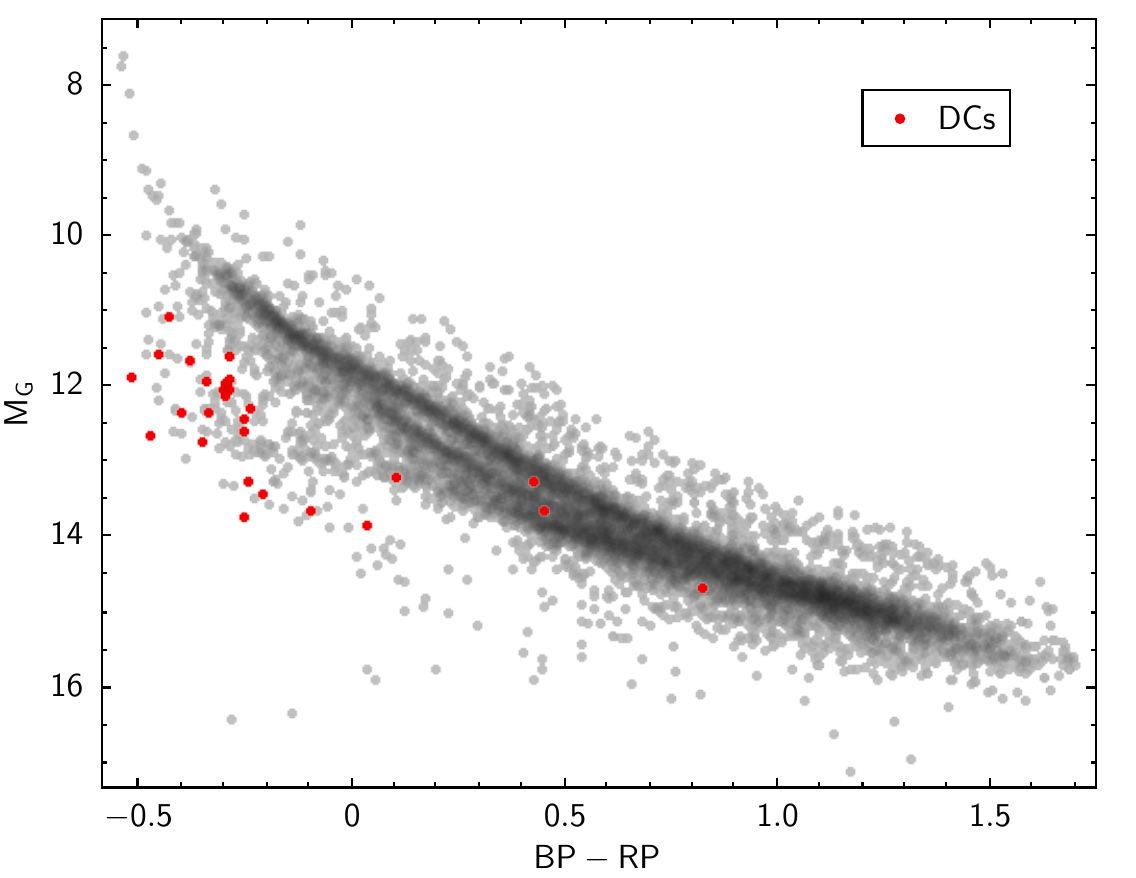}
    \includegraphics[width=0.8\columnwidth,trim=0 0 10 0, clip]{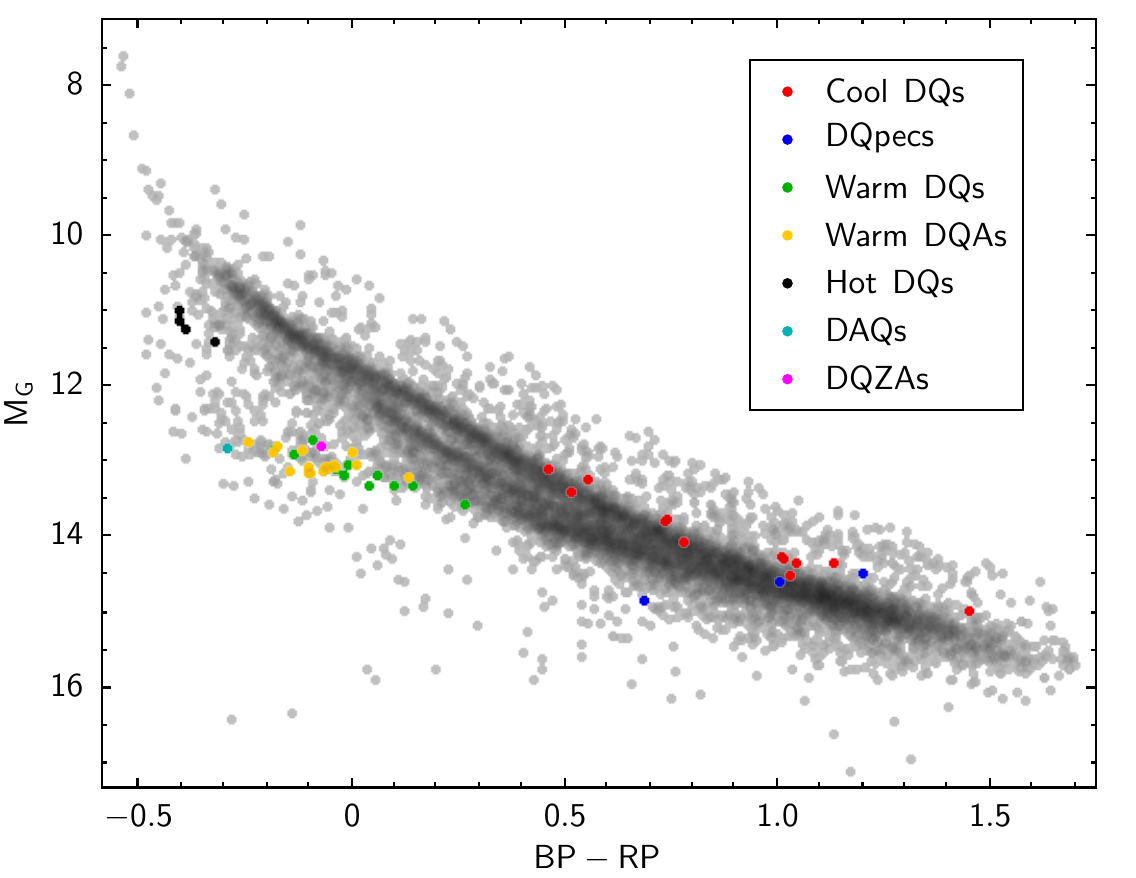}
    \centering
    \includegraphics[width=0.8\columnwidth,trim=0 0 0 0, clip]{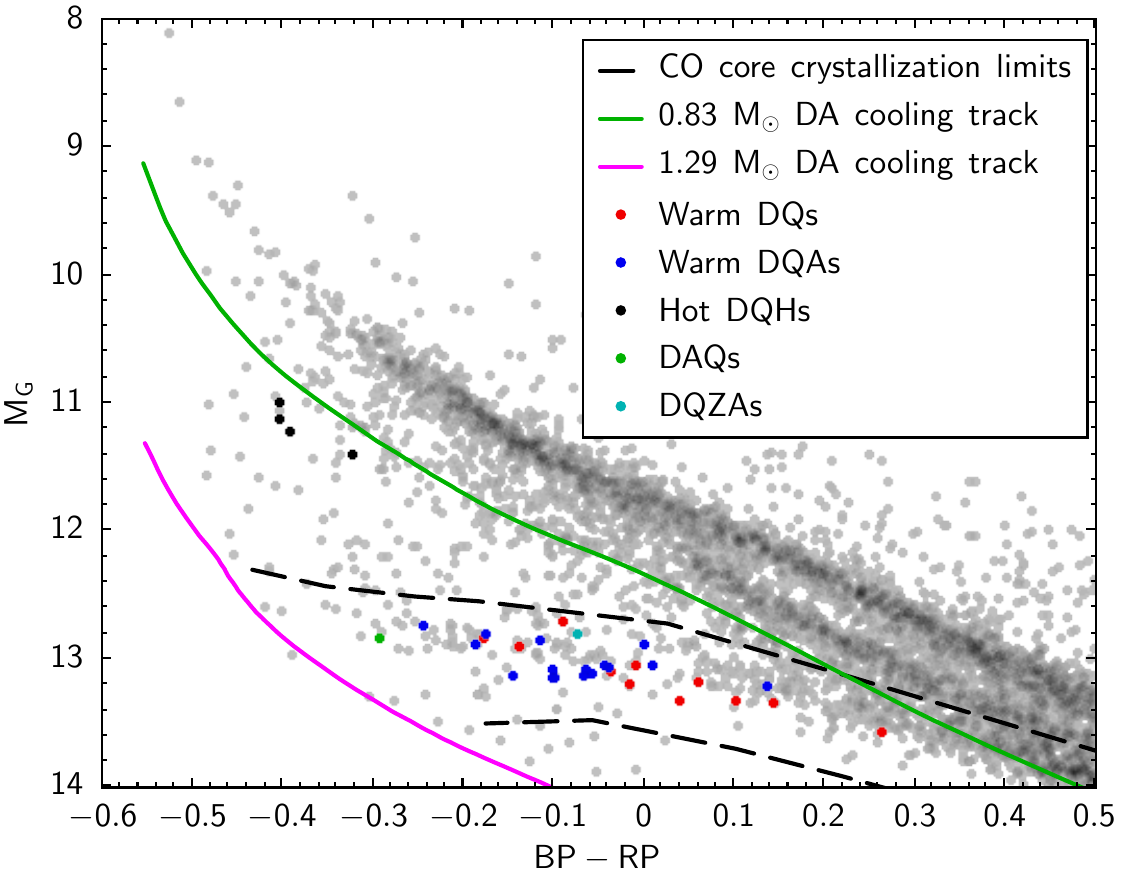}
    \includegraphics[width=0.8\columnwidth,trim=0 0 10 0, clip]{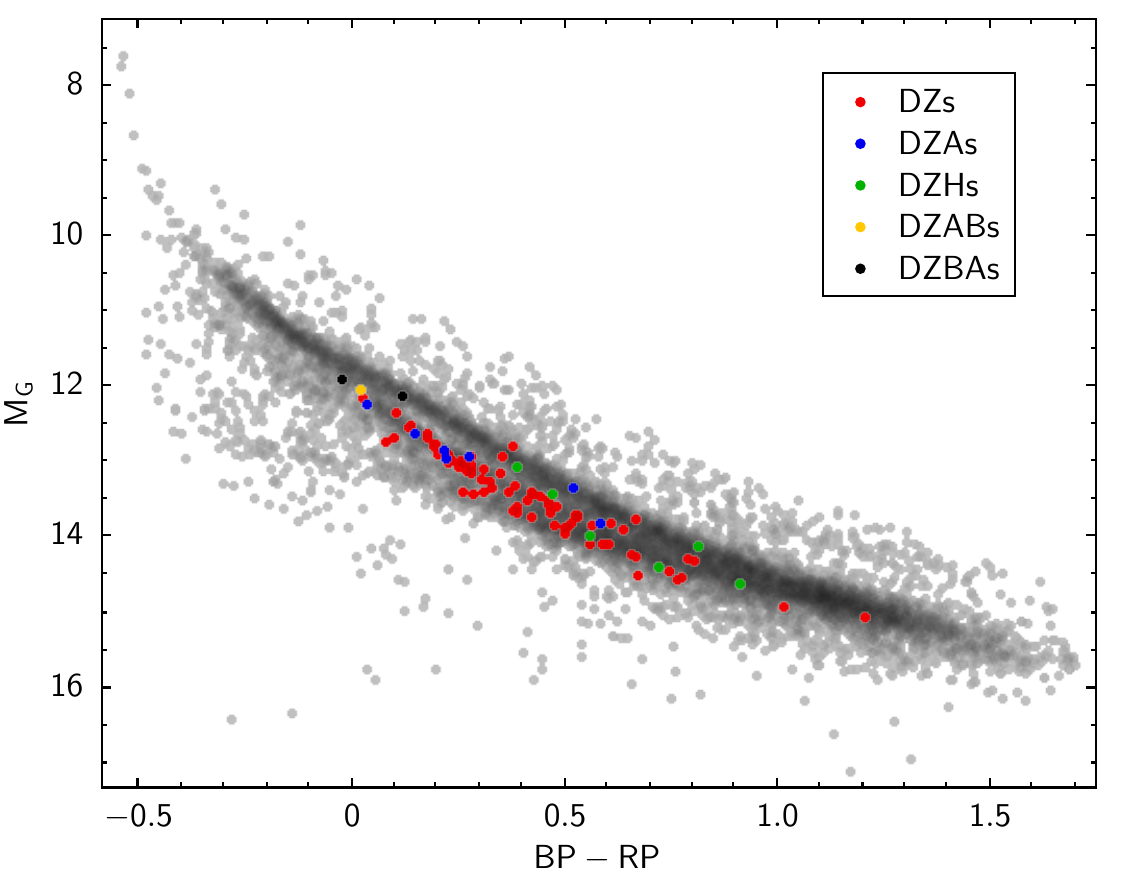}
    \centering
    \includegraphics[width=0.8\columnwidth,trim=0 0 0 0, clip]{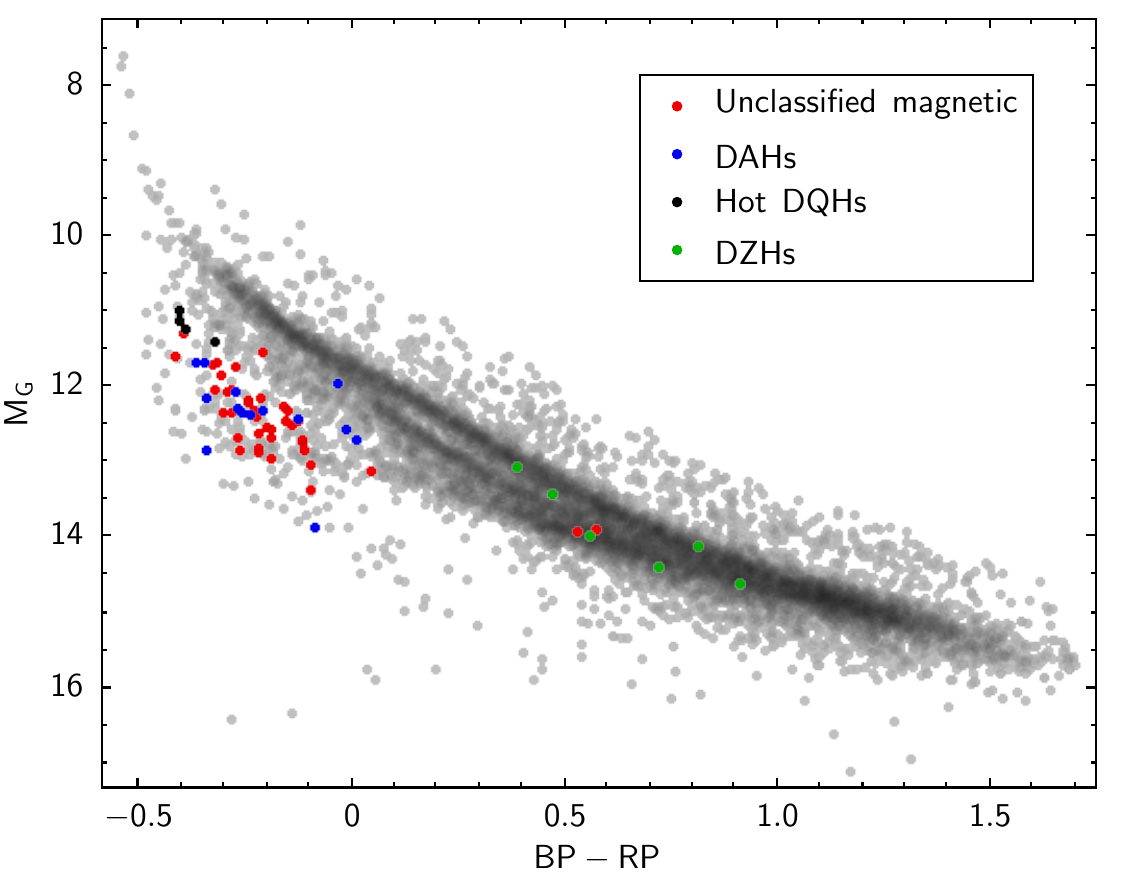}
    \includegraphics[width=0.8\columnwidth,trim=0 0 10 0, clip]{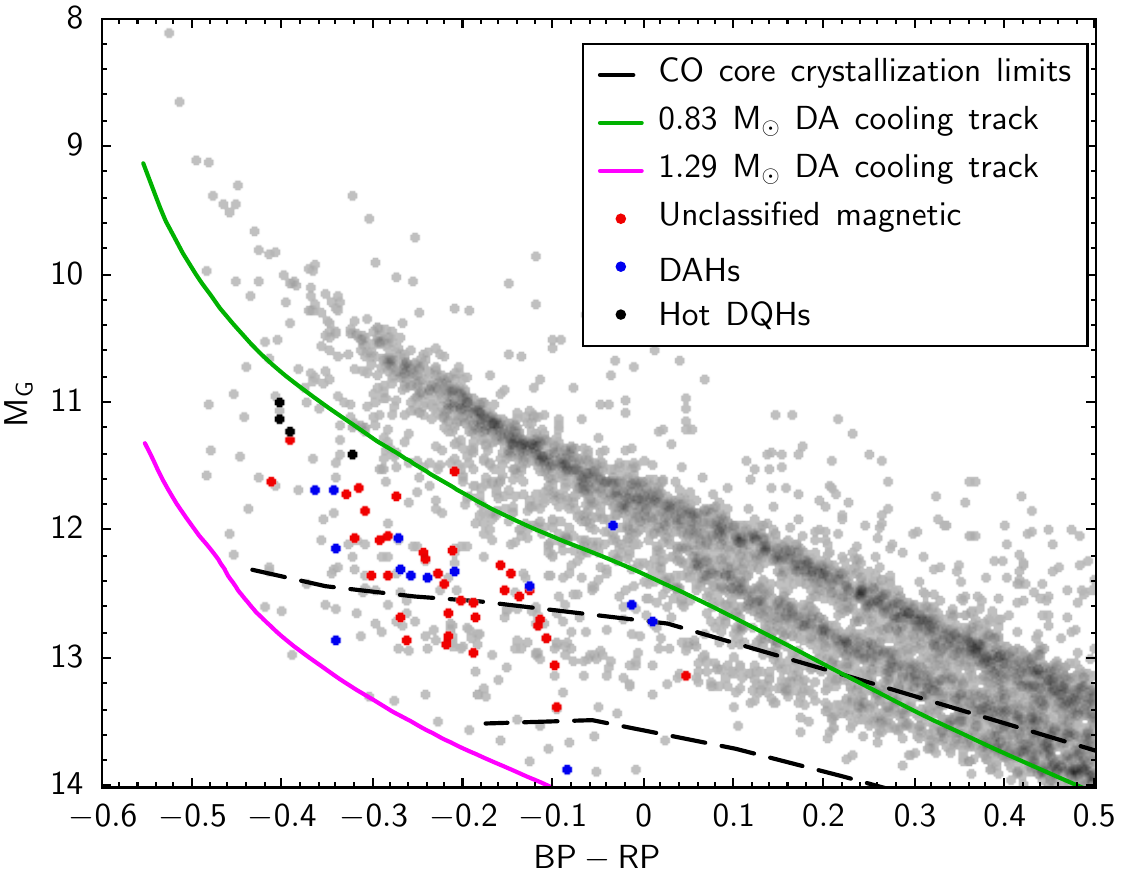}
    \caption{\textit{Gaia} HR diagram for the identified DAs (first row, left panel), DBs and DOs (first row, right panel), DCs (second row, left panel), DQs (second row, right; and third row, left panels), DZs (third row, right panel) and magnetic objects (fourth row, both panels).}
    \label{f:HRs_analysis}
\end{figure*}

\begin{figure*}[h!]
\section{Sample spectra of selected objects}
\label{s:Sample_spectra}
\vspace{1.0cm}

\centering
    \includegraphics[width=1.0\columnwidth,trim=-20 0 0 0, clip]{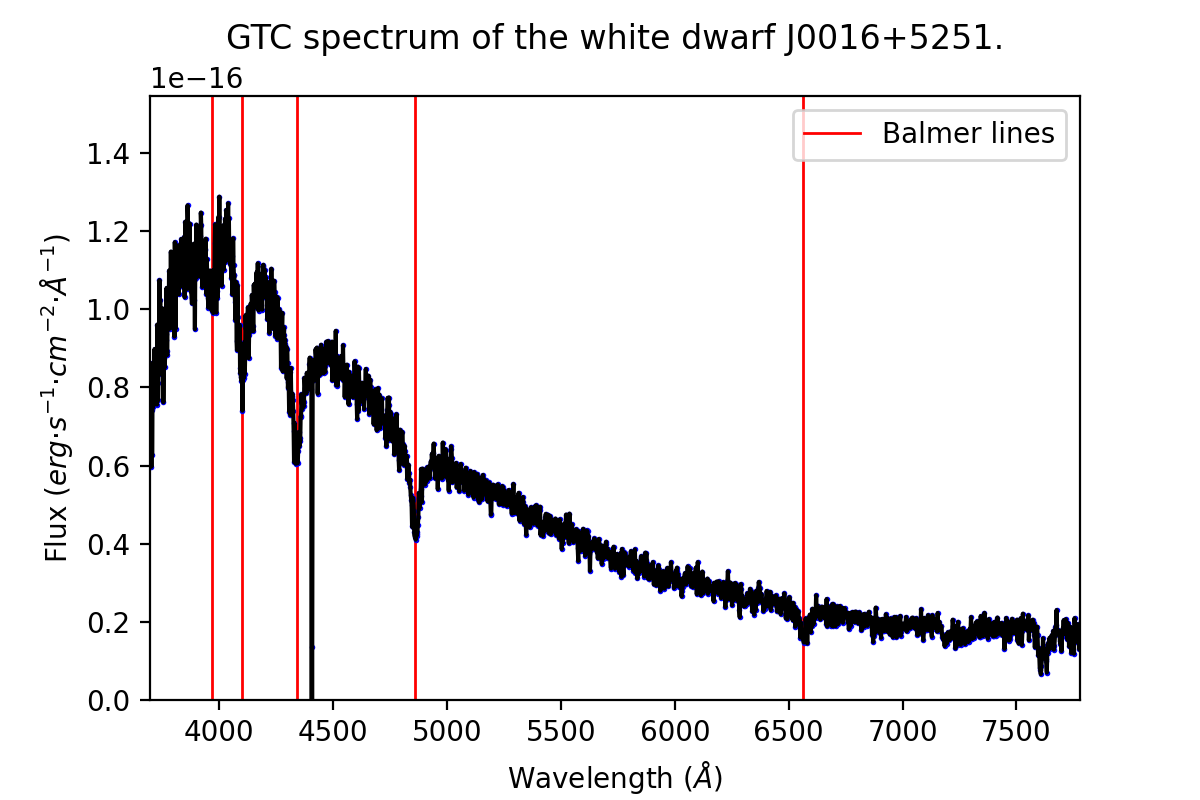}
    \includegraphics[width=1.0\columnwidth,trim=0 0 -20 0, clip]{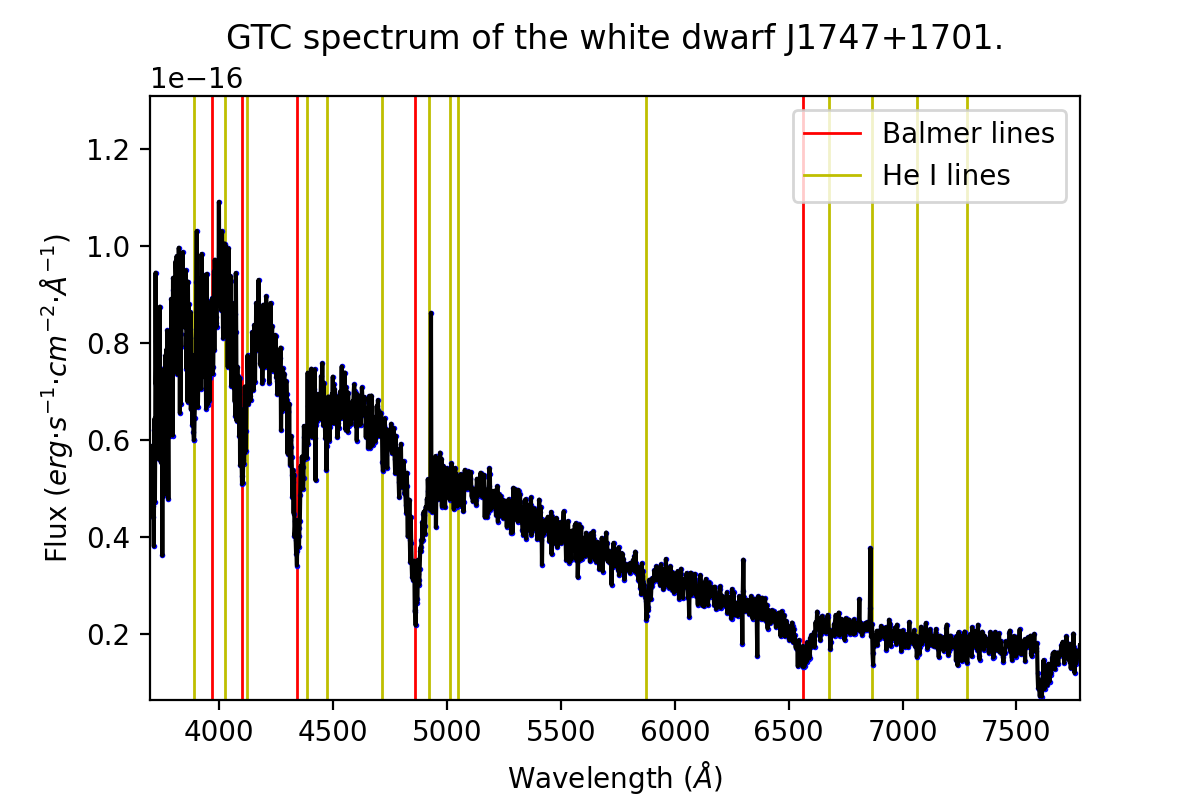}
        \vspace{0.5cm}
    \centering
    \includegraphics[width=1.0\columnwidth,trim=0 0 -20 0, clip]{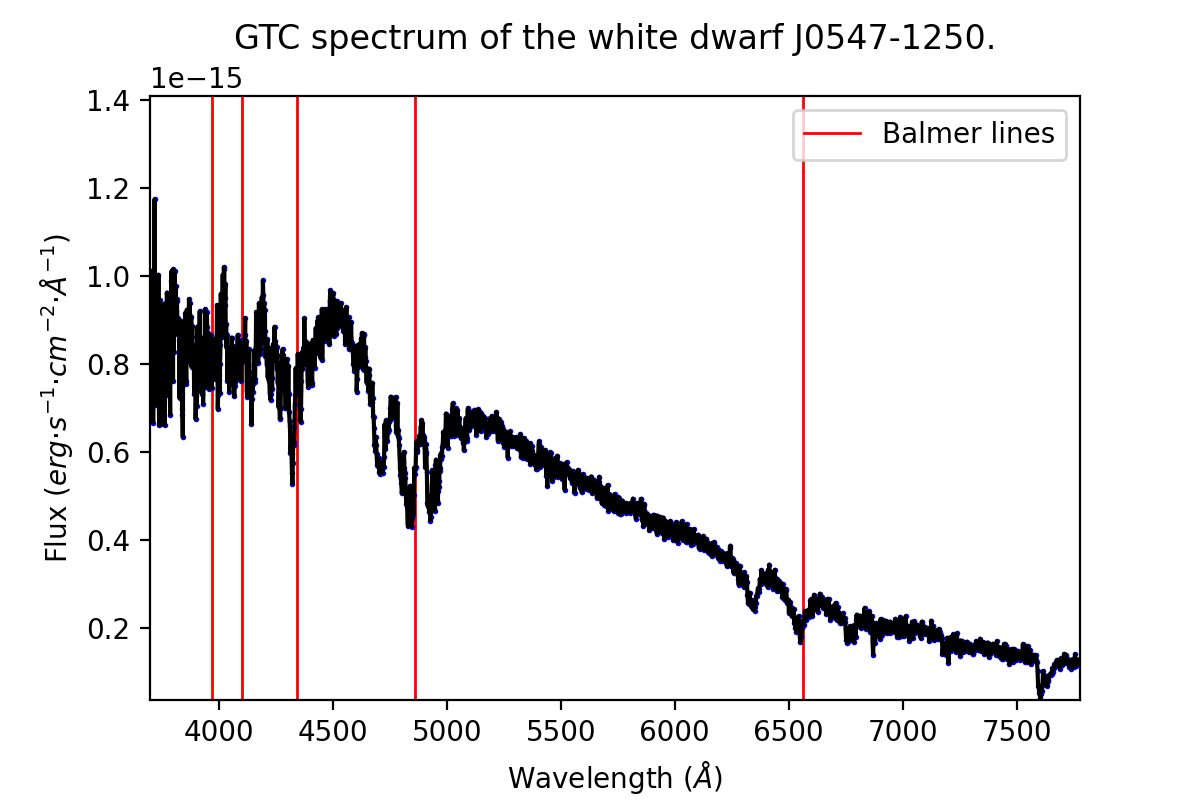}
    \includegraphics[width=1.0\columnwidth,trim=0 0 -20 0, clip]{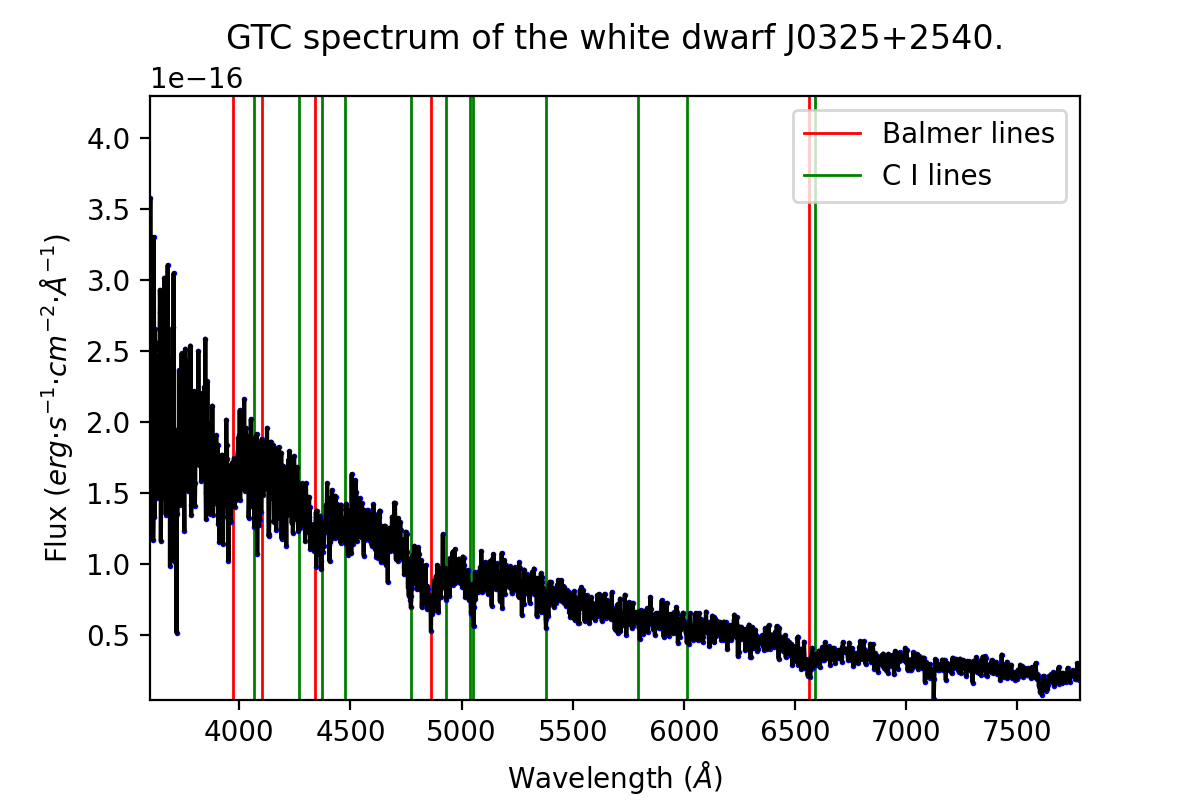}
    \vspace{0.5cm}
    \centering
    \includegraphics[width=1.0\columnwidth,trim=-20 0 0 0, clip]{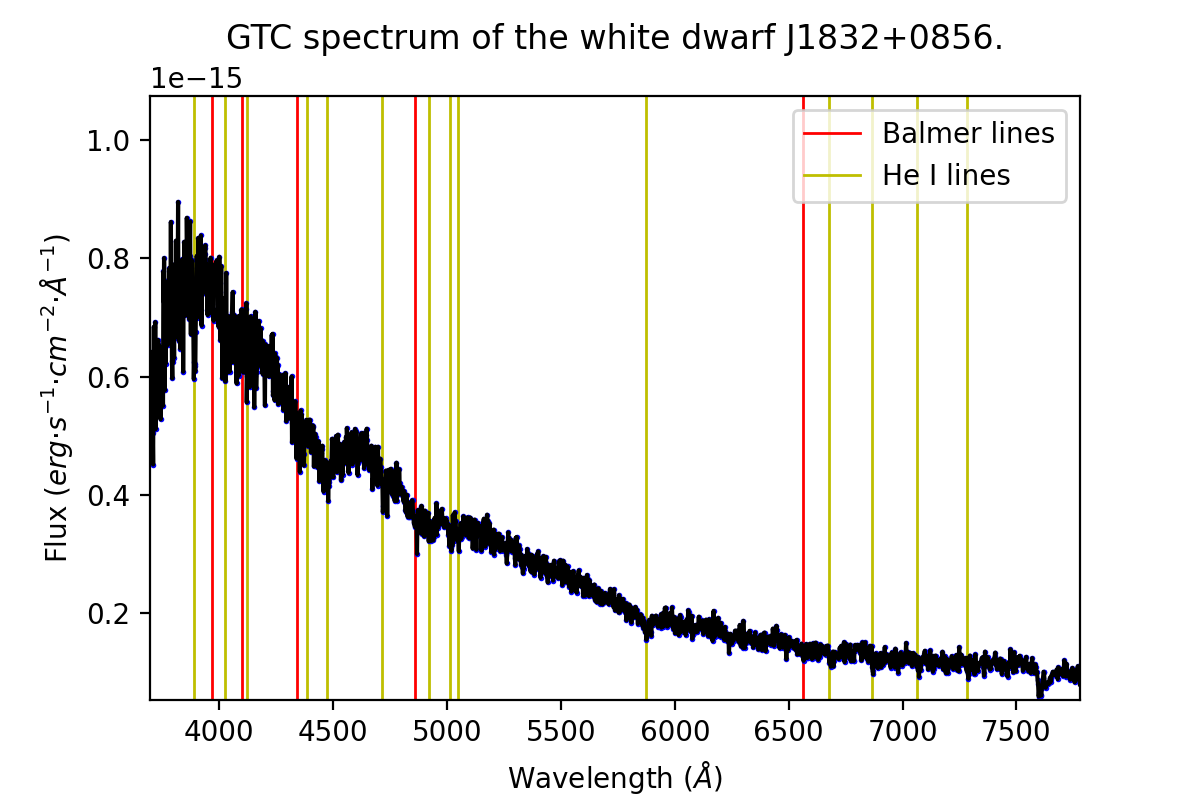}
    \includegraphics[width=1.0\columnwidth,trim=0 0 -20 0, clip]{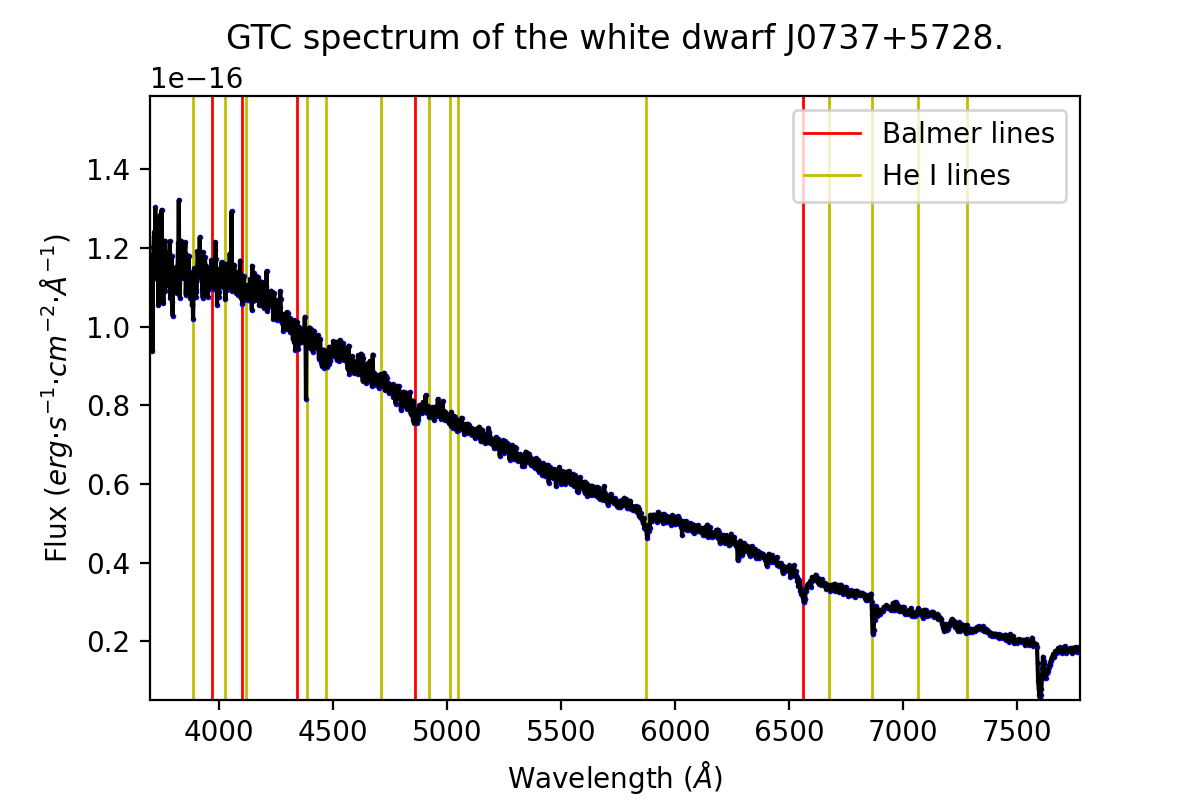}
    \caption{GTC spectra of a pure DA (first panel), a DAB (second panel), a DAH (third panel), a DAQ (fourth panel), a pure DB (fifth panel) and a DBA (sixth panel).}
    \label{f:Sp1}
        \vspace{0.5cm}
\end{figure*}

\begin{figure*}[h!]

    \centering
    \includegraphics[width=1.0\columnwidth,trim=0 0 -20 0, clip]{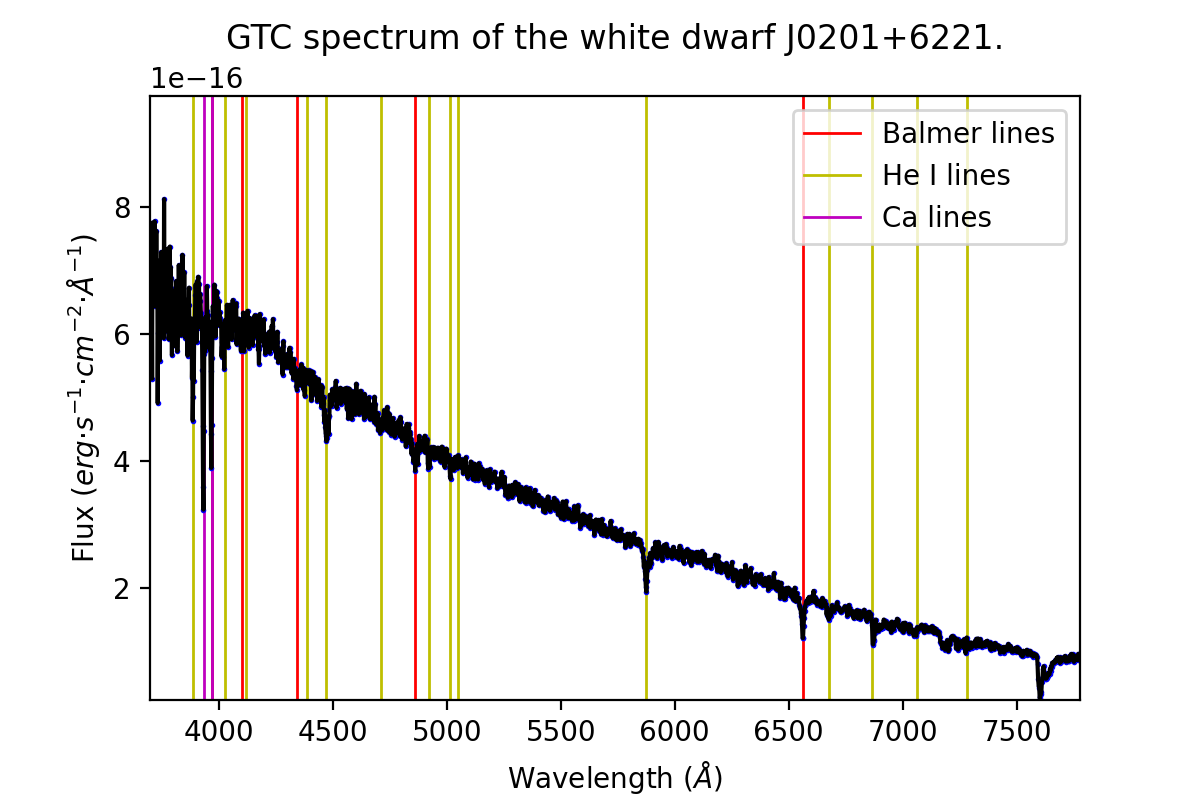}
    \includegraphics[width=1.0\columnwidth,trim=-20 0 0 0, clip]{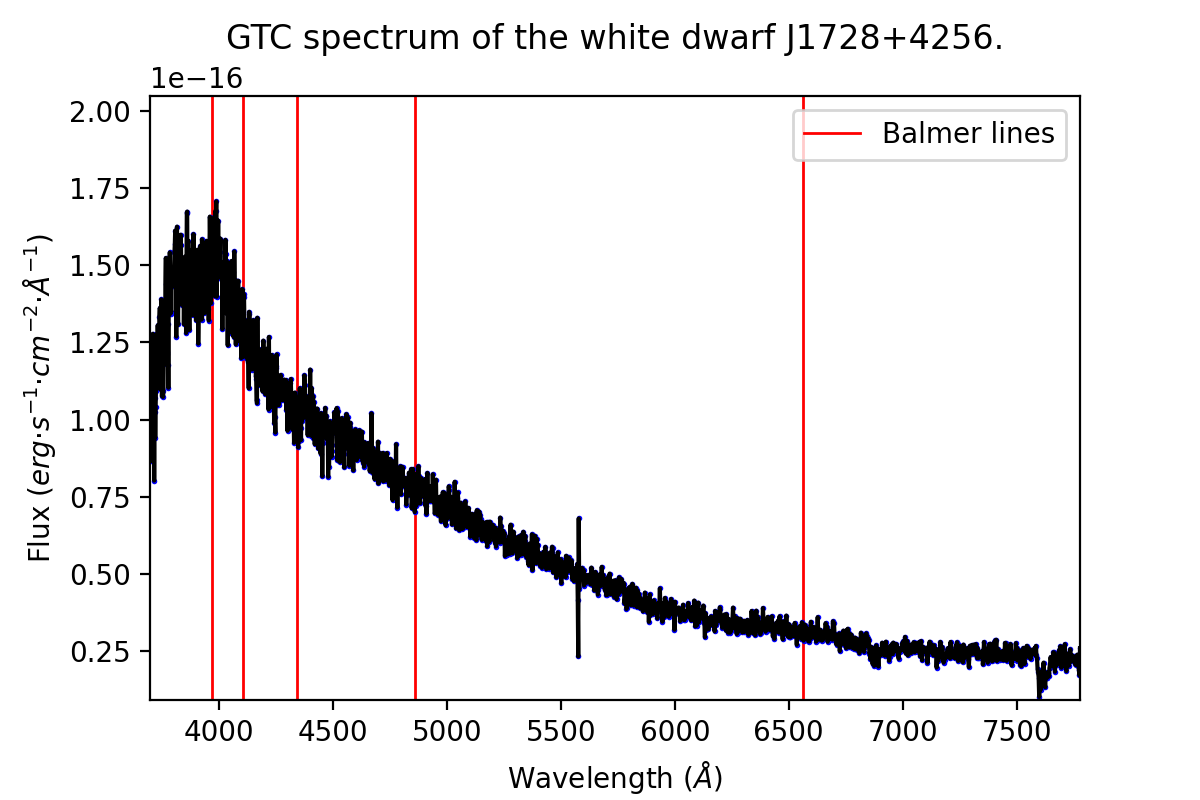}
    \centering
    \includegraphics[width=1.0\columnwidth,trim=0 0 -20 0, clip]{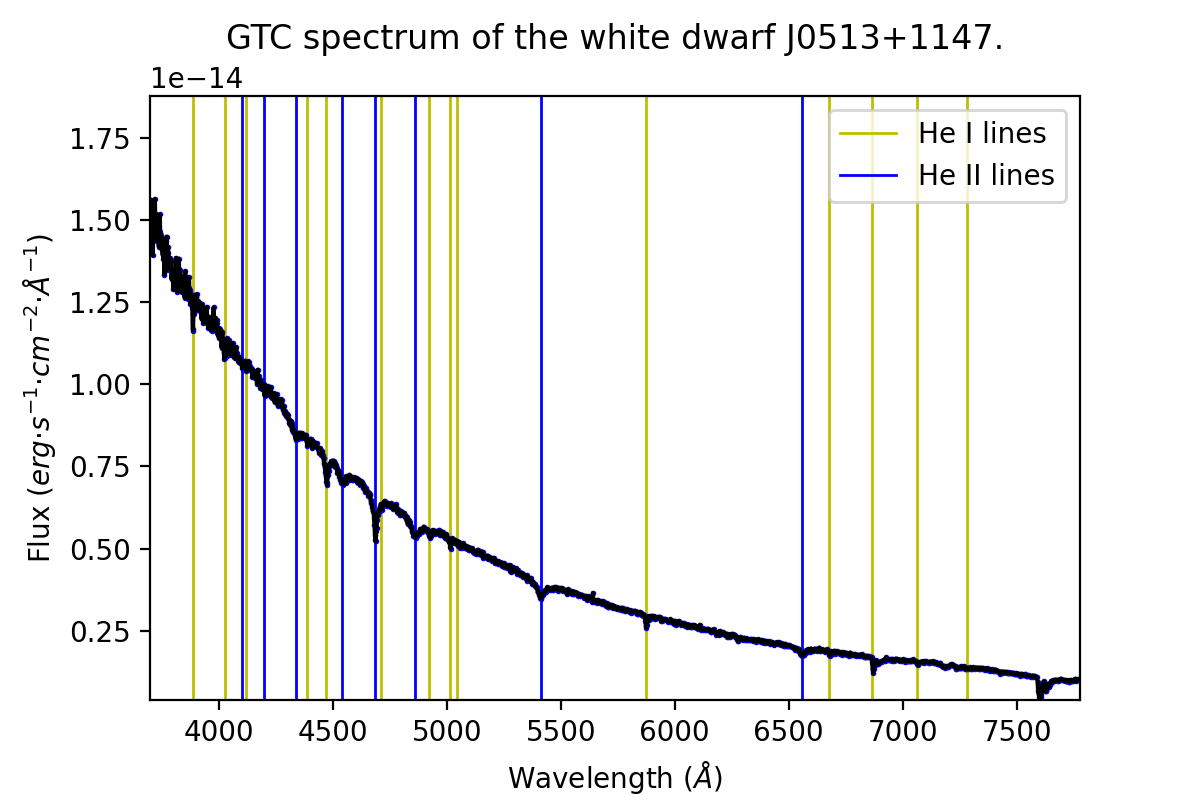}
    \includegraphics[width=1.0\columnwidth,trim=-20 0 0 0, clip]{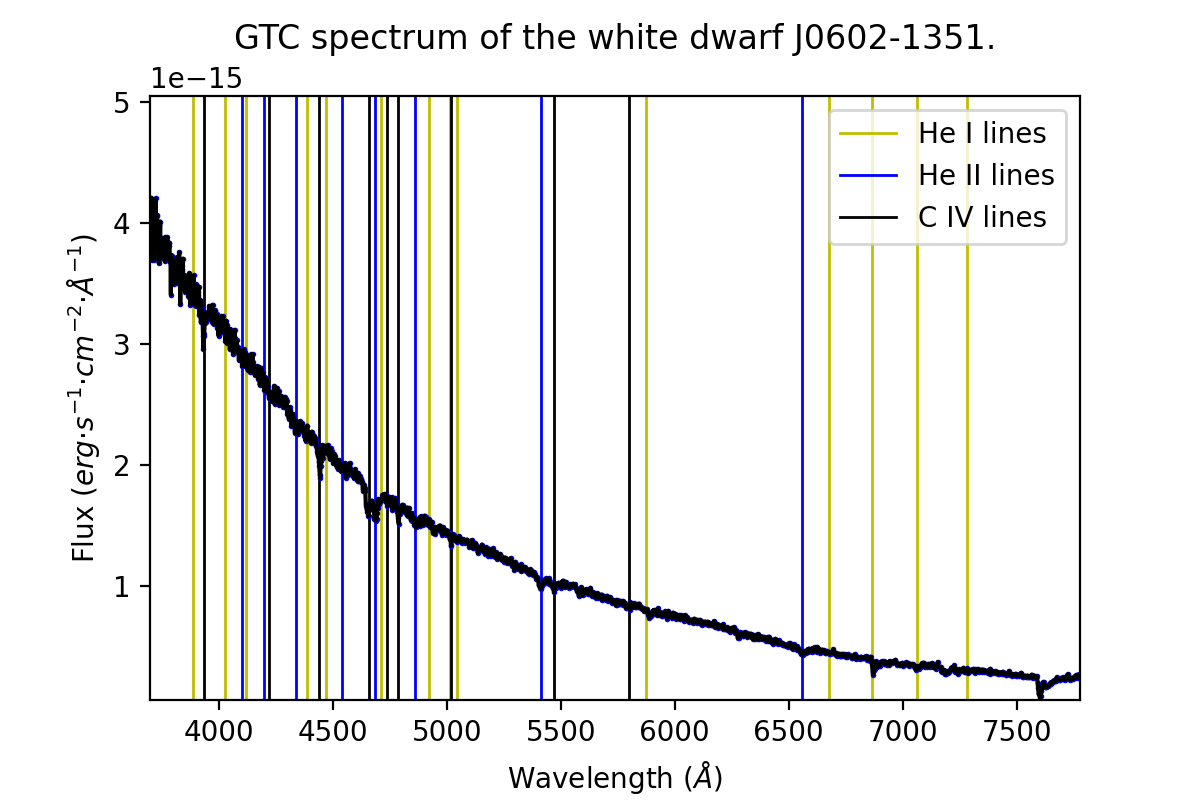}
    \vspace{0.5cm}
    \centering
    \includegraphics[width=1.0\columnwidth,trim=-20 0 0 0, clip]{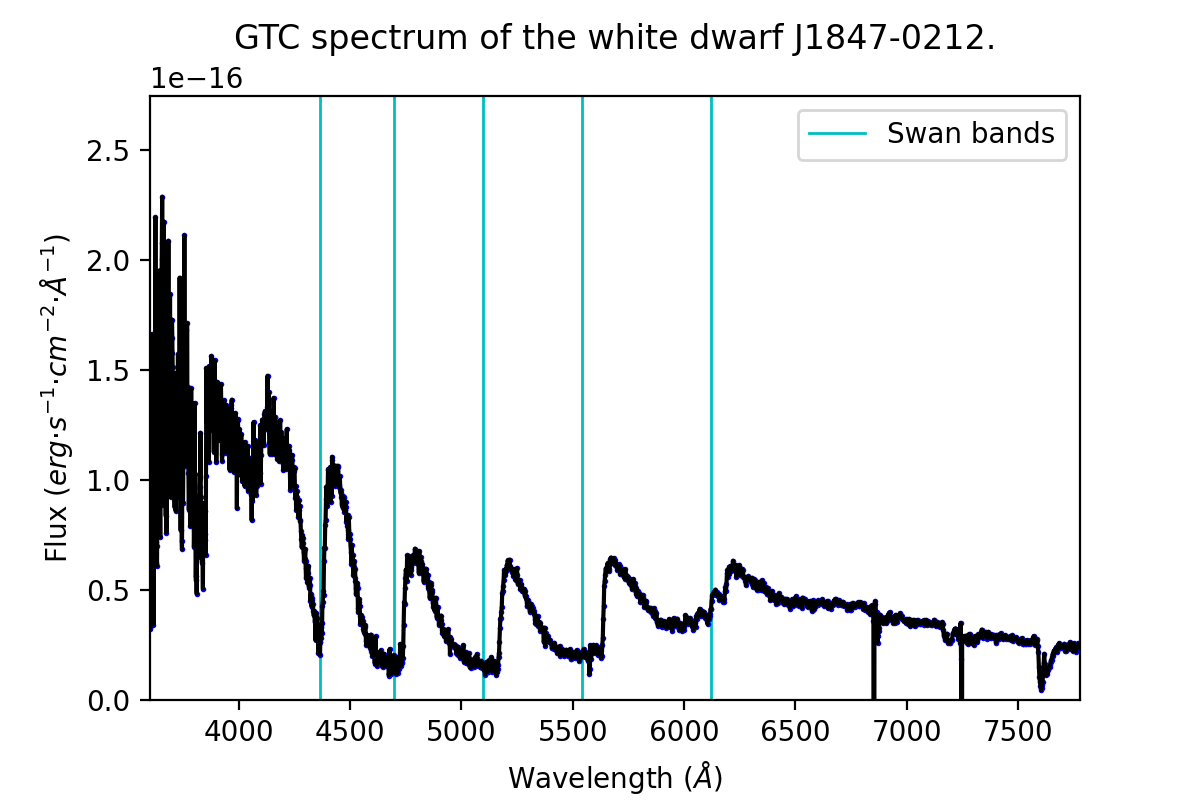}
    \includegraphics[width=1.0\columnwidth,trim=0 0 -20 0, clip]{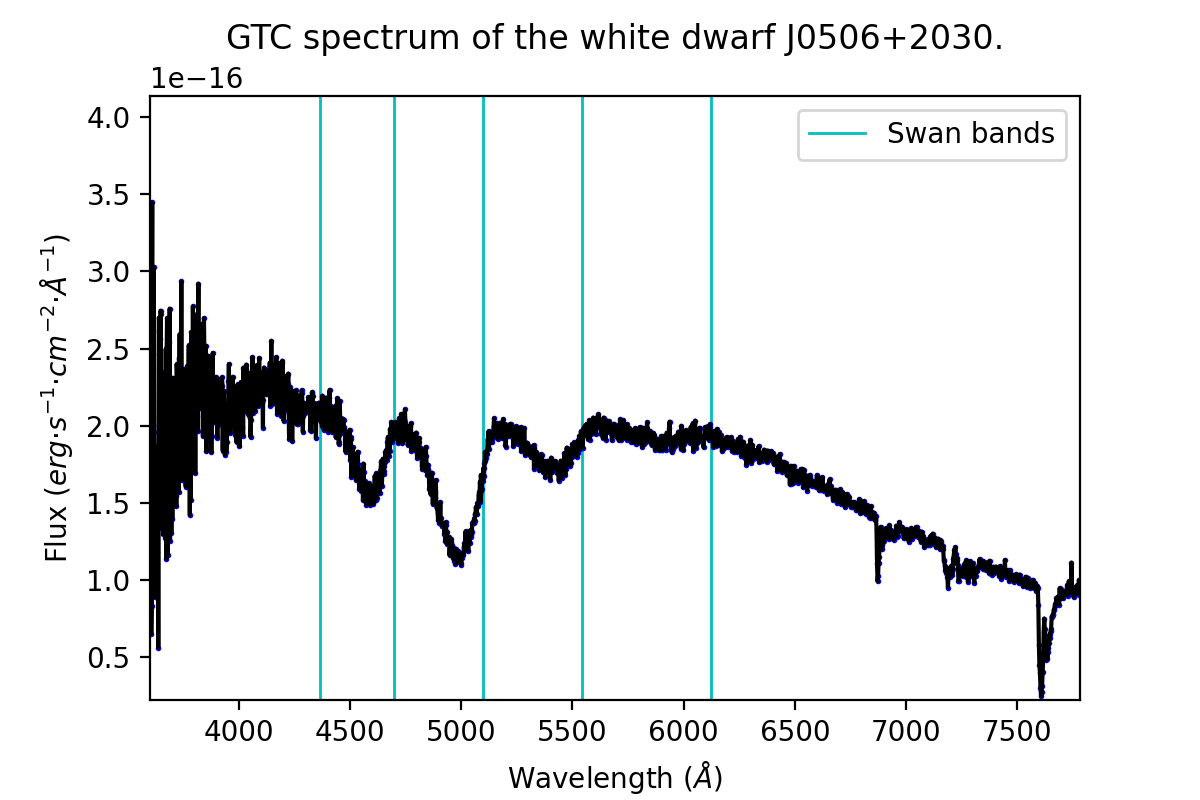}
    \vspace{0.5cm}
    \caption{GTC spectra of a DBAZ (first panel), a DC (second panel), a pure DO (third panel), a DOZ (fourth panel), a cool DQ (fifth panel) and a DQpec (sixth panel).}
    \label{f:Sp2}
\end{figure*}

\begin{figure*}[h!]
\vspace{1.0cm}

\centering
    \includegraphics[width=1.0\columnwidth,trim=0 0 -20 0, clip]{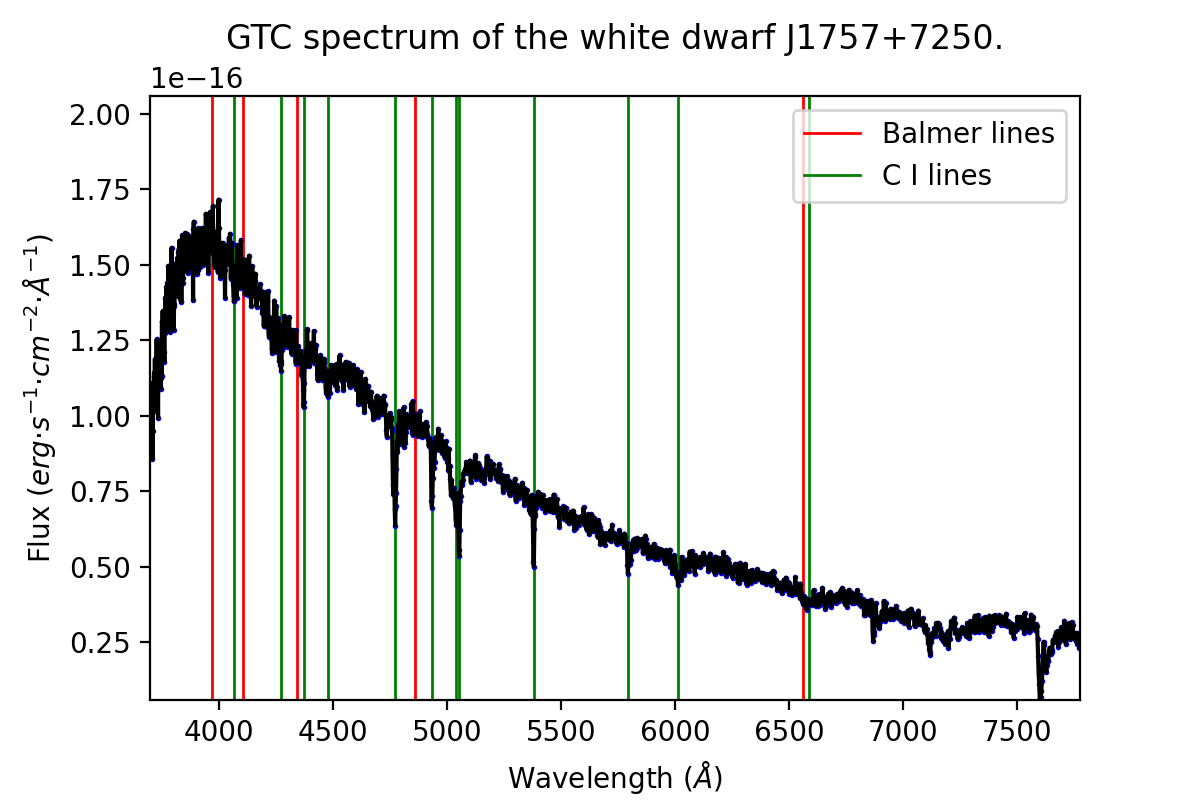}
    \includegraphics[width=1.0\columnwidth,trim=-20 0 0 0, clip]{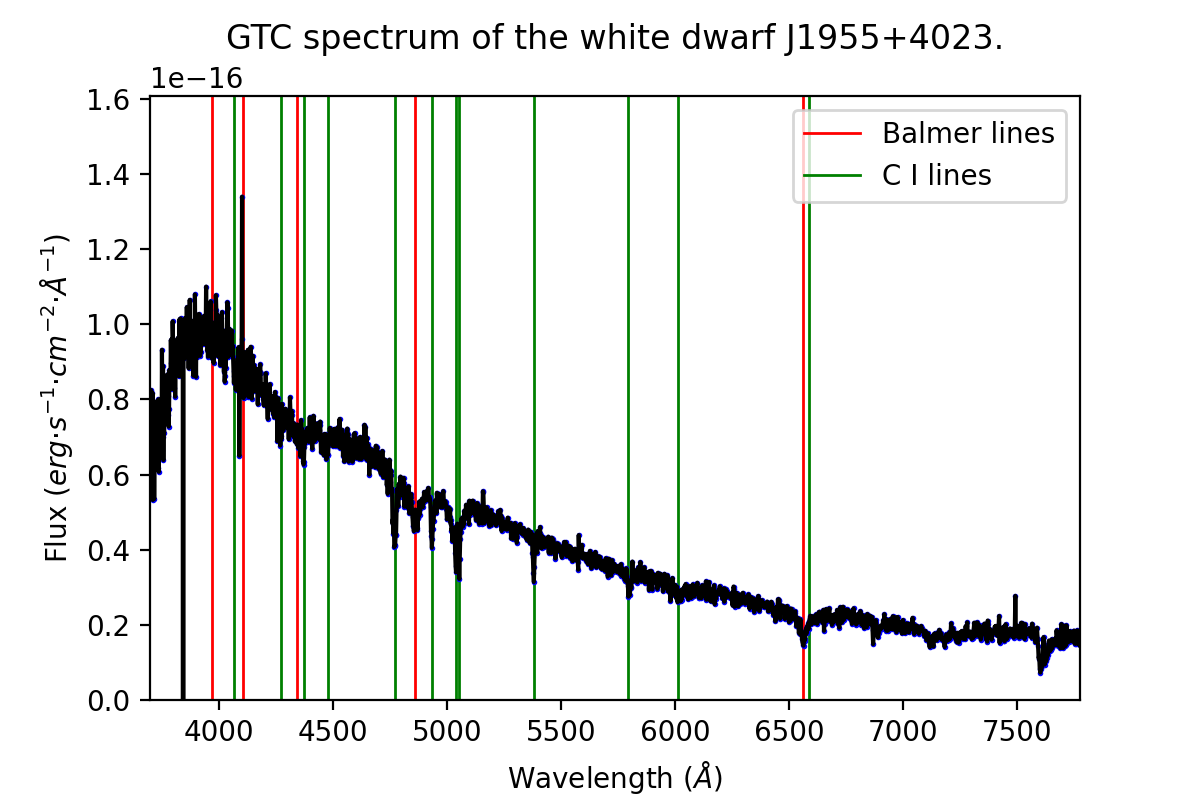}
    \vspace{0.5cm}
    \centering
    \includegraphics[width=1.0\columnwidth,trim=0 0 -20 0, clip]{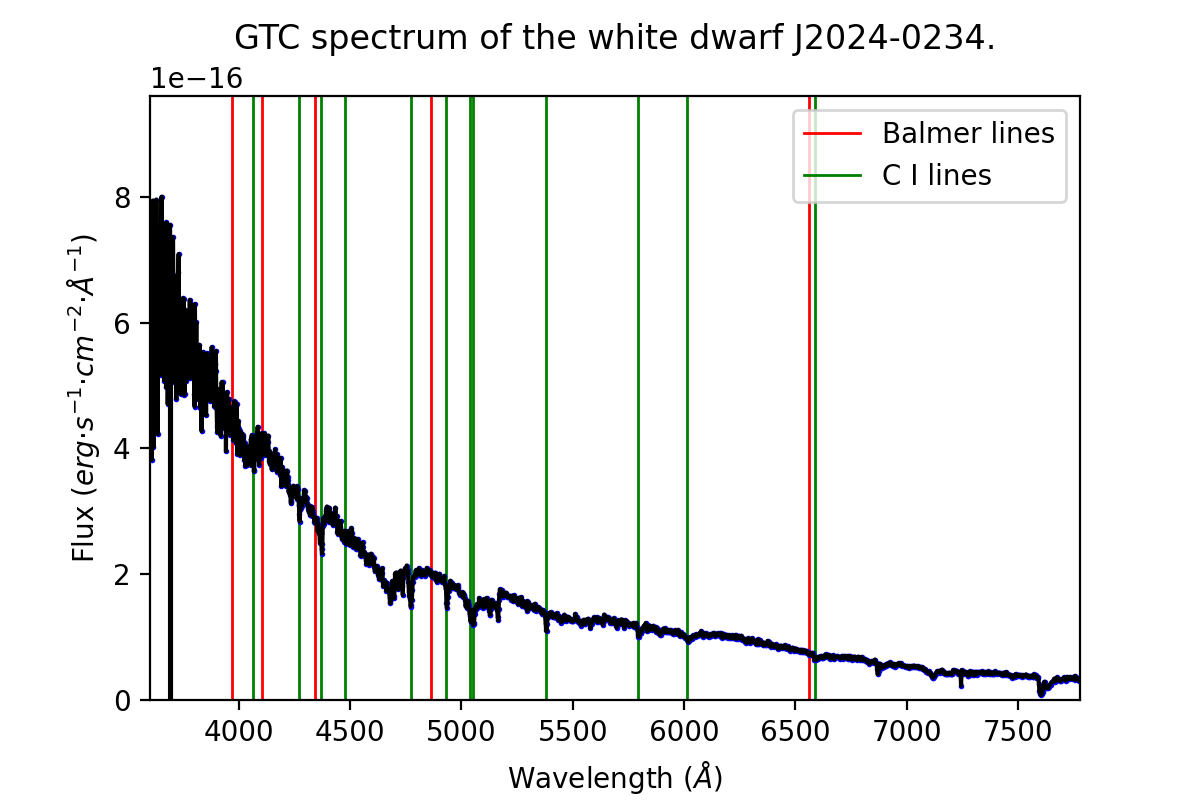}
    \includegraphics[width=1.0\columnwidth,trim=-20 0 0 0, clip]{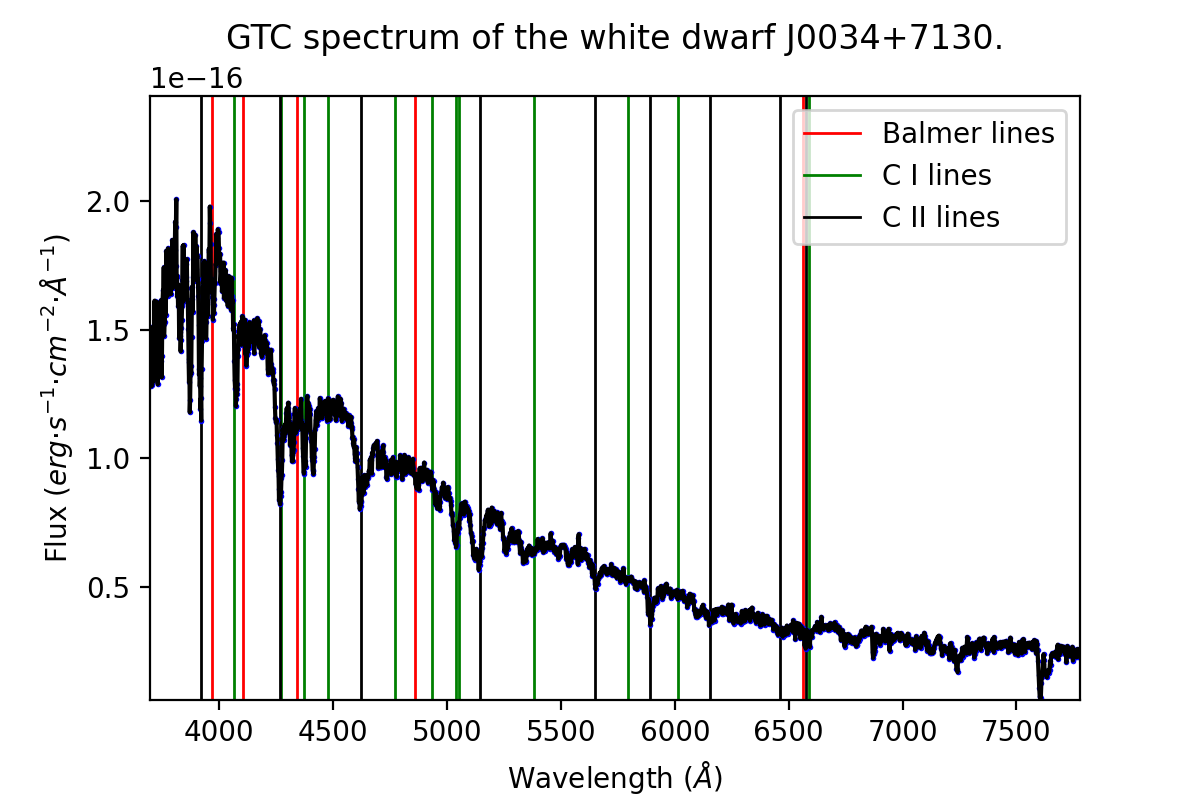}
    \vspace{0.5cm}
    \centering
    \includegraphics[width=1.0\columnwidth,trim=0 0 -20 0, clip]{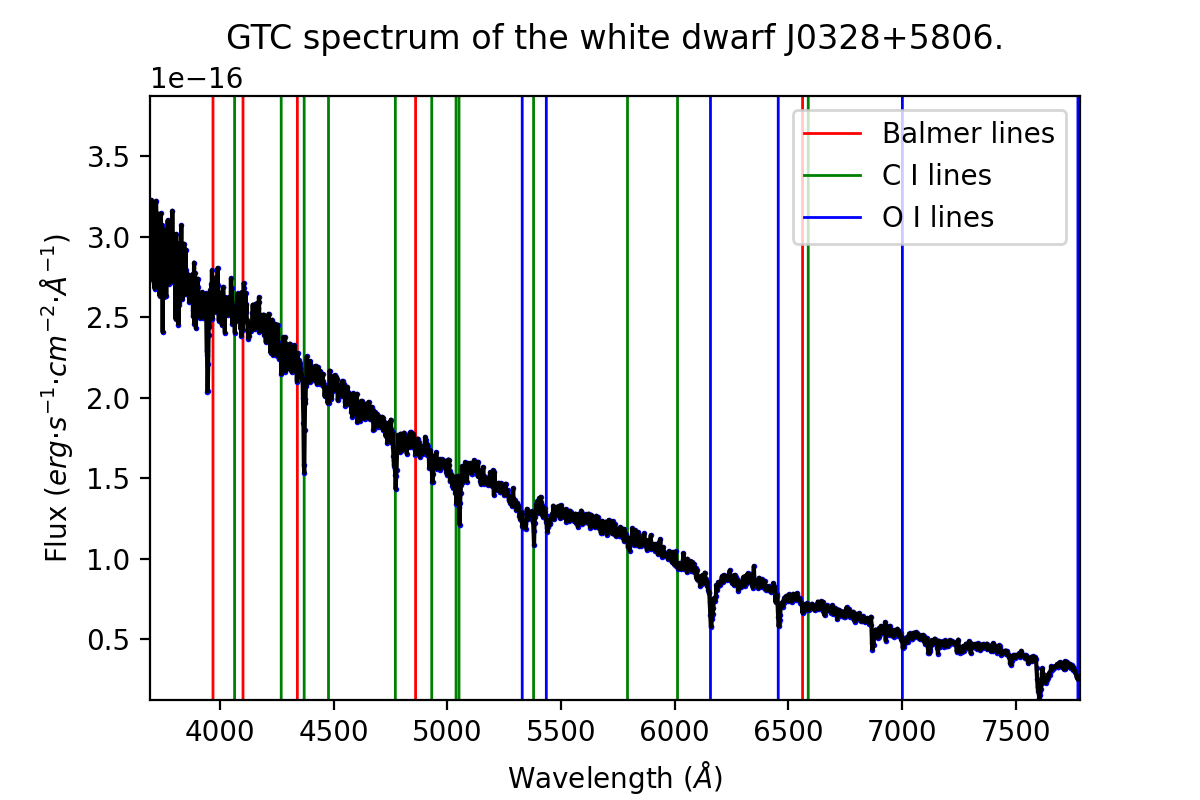}
    \includegraphics[width=1.0\columnwidth,trim=-20 0 0 0, clip]{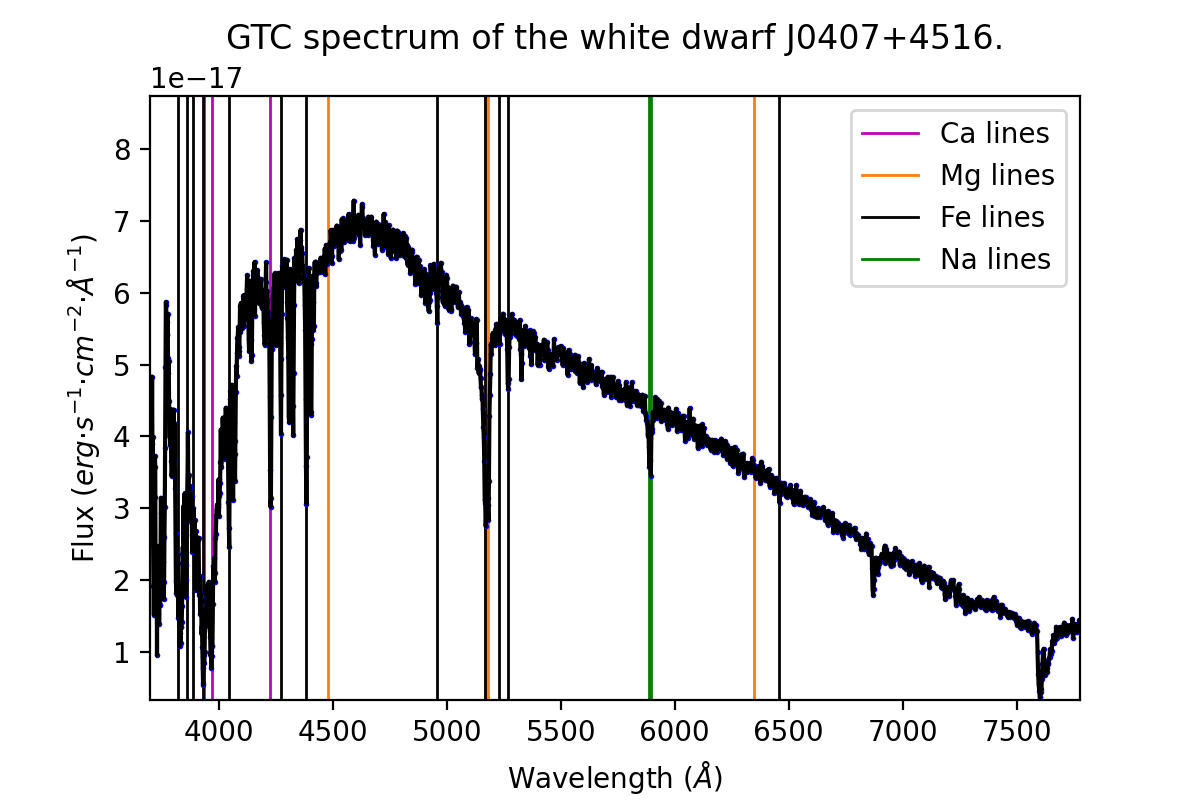}
    
    \caption{GTC spectra of a warm DQ (first panel), a warm DQA (second panel), a warm DQ displaying both weak C$_{2}$ Swan bands and \ion{C}{I} spectral lines (third panel), a hot DQ (fourth panel), a warm DQZA (fifth panel) and a pure DZ (sixth panel).}
    \label{f:Sp3}
        \vspace{0.5cm}
\end{figure*}

\begin{figure*}[h!]
\centering
    \includegraphics[width=1.0\columnwidth,trim=-20 0 0 0, clip]{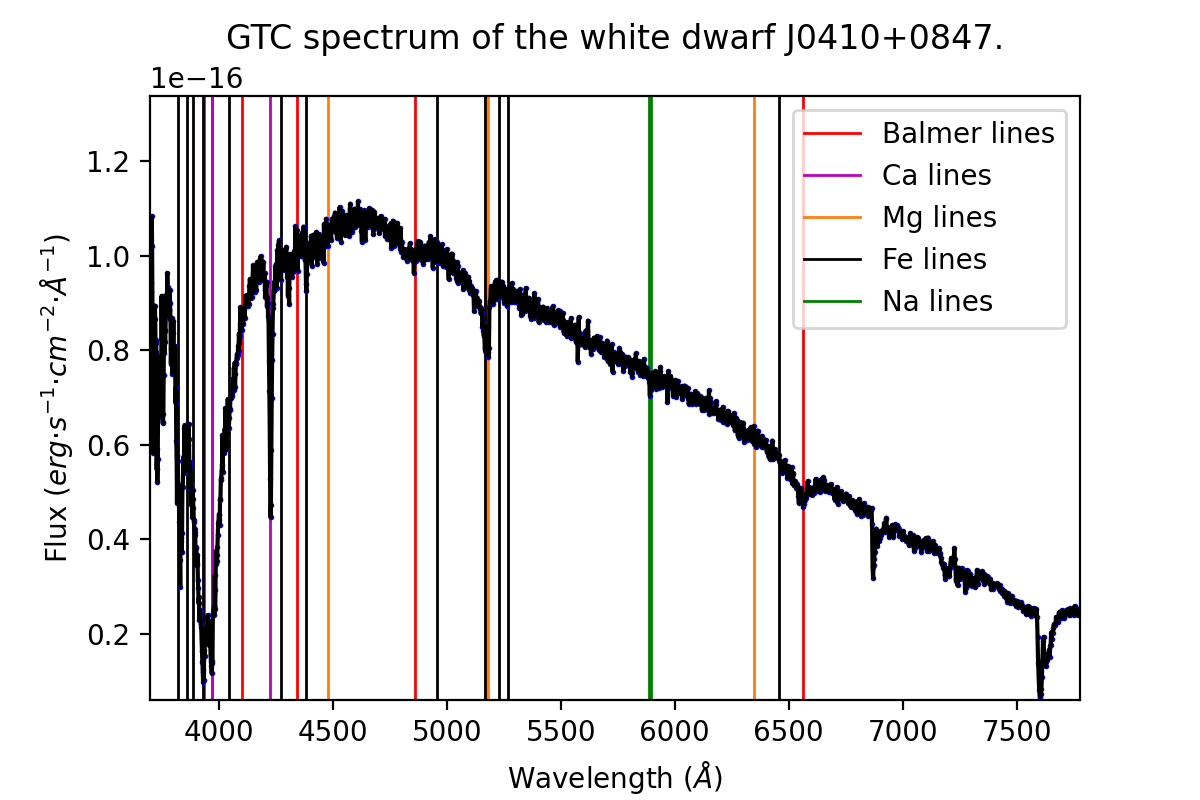}
    \includegraphics[width=1.0\columnwidth,trim=0 0 -20 0, clip]{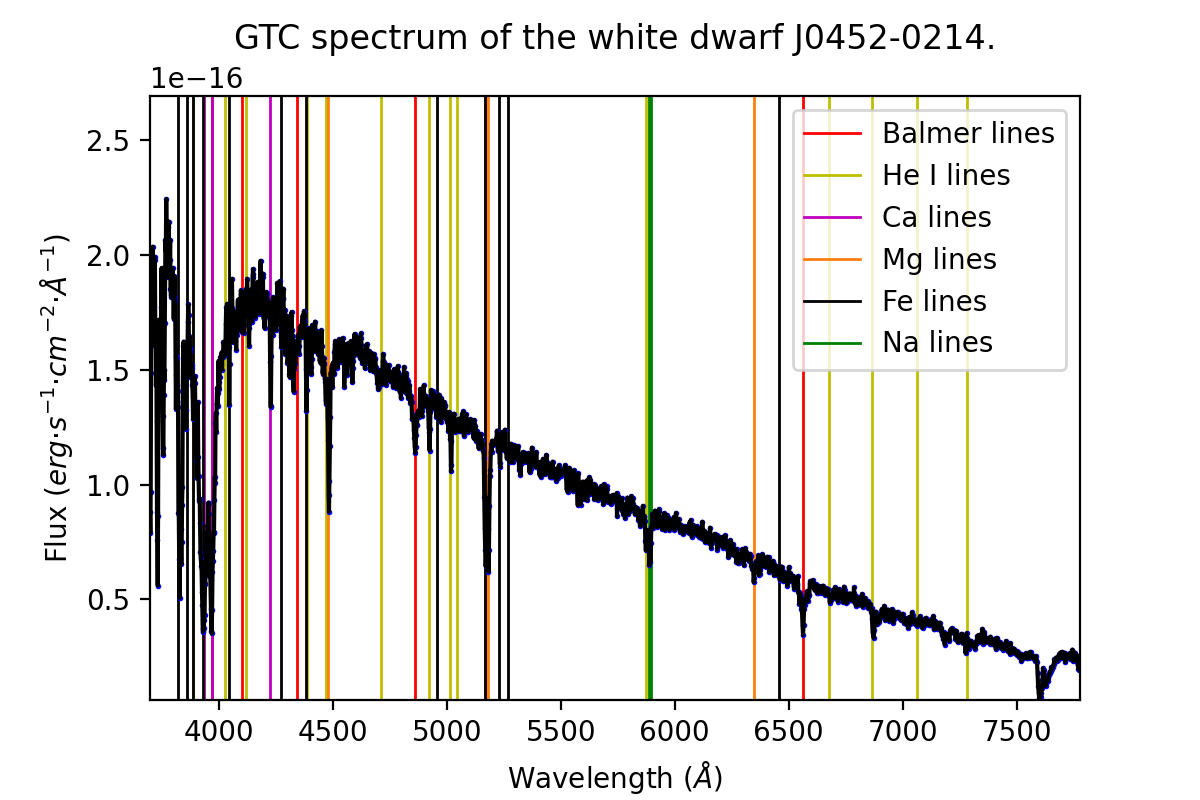}
        \vspace{0.5cm}
    \centering
    \includegraphics[width=1.0\columnwidth,trim=-20 0 0 0, clip]{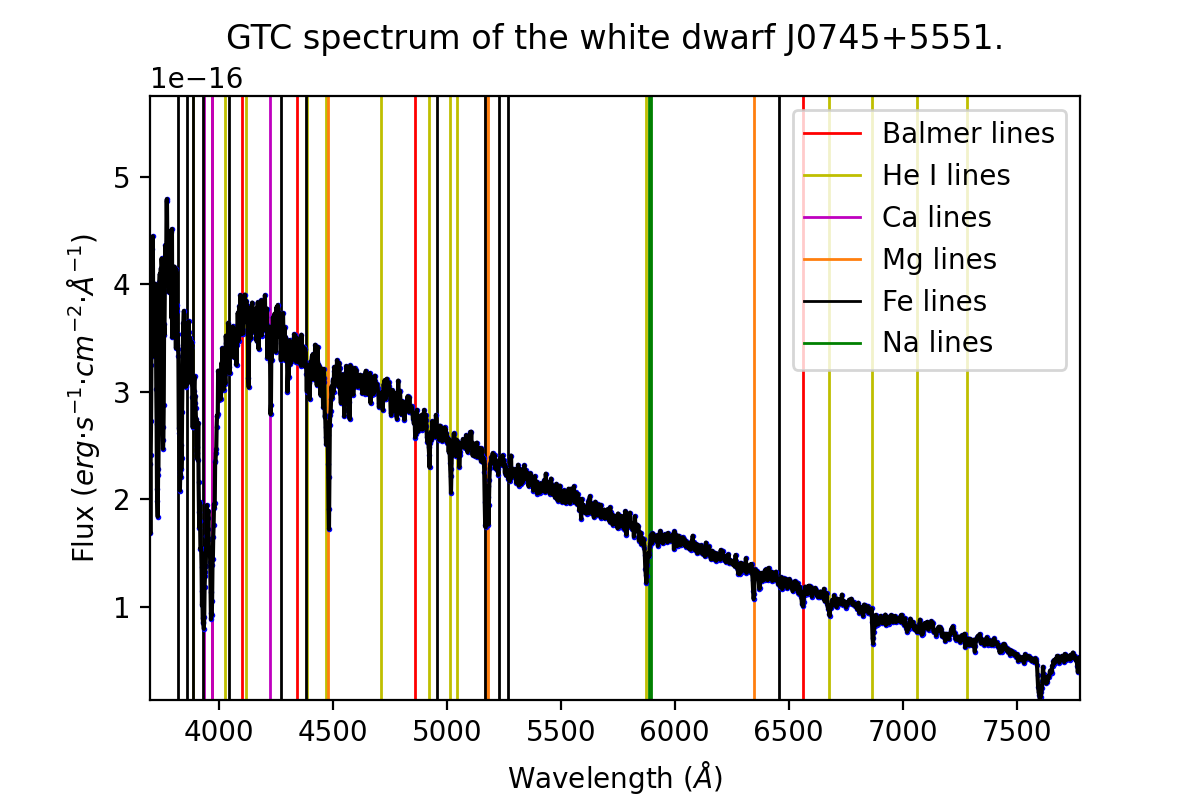}
    \includegraphics[width=1.0\columnwidth,trim=0 0 -20 0, clip]{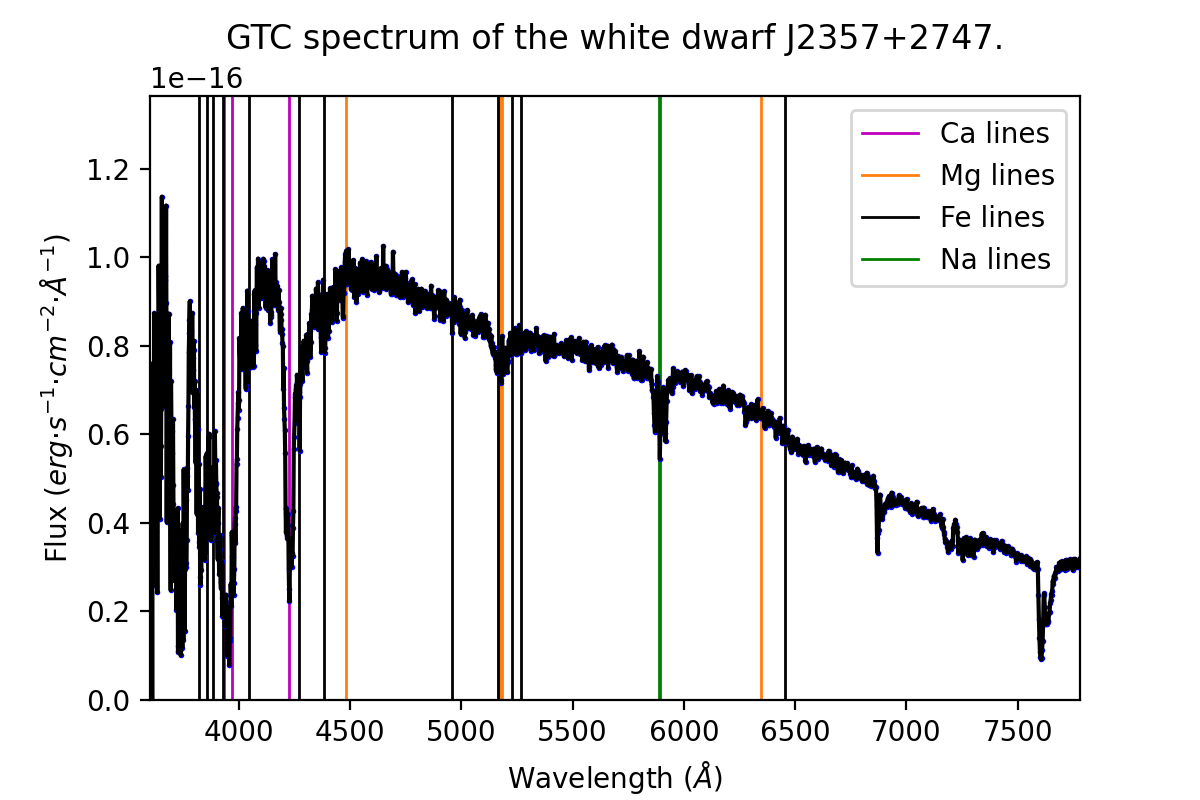}
        \vspace{0.5cm}
    \centering
    \includegraphics[width=1.0\columnwidth,trim=-20 0 0 0, clip]{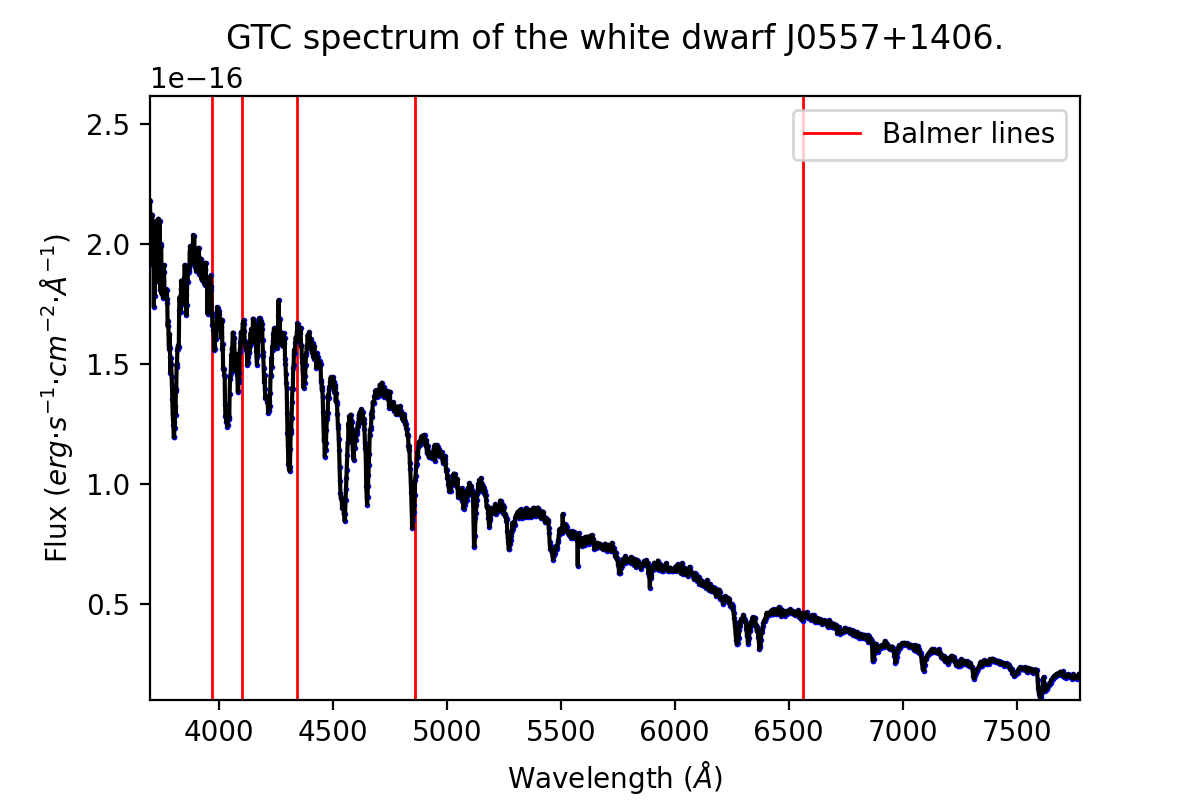}
    \includegraphics[width=1.0\columnwidth,trim=0 0 -20 0, clip]{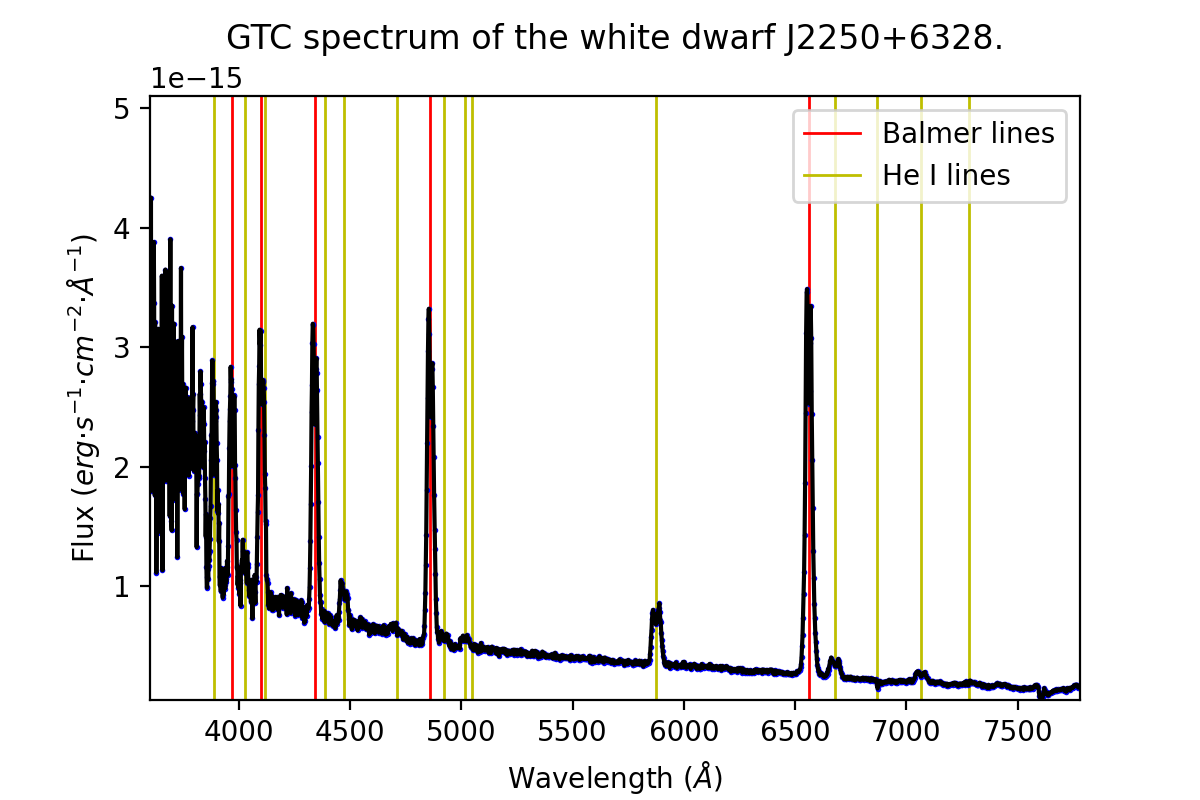}
    
    \caption{GTC spectra of a DZA (first panel), a DZAB (second panel), a DZBZ (third panel), a DZH (fourth panel), an unclassified magnetic white dwarf with clear signs of Zeeman splitting (fifth panel) and a CV (sixth panel).}
    \label{f:Sp4}
        \vspace{0.5cm}
\end{figure*}

\end{appendix}
\end{document}